\documentclass[12pt]{article}
\usepackage{titlesec}
\usepackage{titling}
\usepackage{amsmath, amsthm, epsfig, amssymb}
\usepackage{hyperref}
\usepackage[font={small,sf},labelfont={small,sf,bf}]{caption}
\usepackage[normalem]{ulem}

\usepackage{authblk}
\usepackage{verbatim}
\usepackage[titles]{tocloft}

\usepackage{epigraph}
\usepackage[utf8]{inputenc}
\usepackage{cleveref}
\crefname{section}{§}{§§}

\renewcommand{\th}{\theta}

\usepackage{float}

\usepackage{enumitem}
\setlist[itemize]{leftmargin=*}
\setlist[itemize]{itemsep=-1ex}
\setlist[itemize]{topsep=0ex}

\title{{\normalfont\Large Explanatory models in neuroscience: \vspace{-5pt}
\linebreak Part 1 -- taking mechanistic abstraction seriously
}\vspace{-15pt}
}

\author{Rosa Cao and Daniel Yamins\thanks{Department of Philosophy (RC); Departments of Psychology \& Computer Science (DY); and Wu Tsai Neurosciences Institute (RC \& DY), Stanford University.}\vspace{-75pt}}

\date{}

\begin{document}

\maketitle

\setlength{\cftbeforesecskip}{3pt}
\setcounter{tocdepth}{1}
\tableofcontents

\thispagestyle{empty}


\begin{abstract}
Despite the recent success of neural network models in mimicking animal performance on visual perceptual tasks, critics worry that these models fail to illuminate brain function. We take it that a central approach to explanation in systems neuroscience is that of mechanistic modeling, where understanding the system is taken to require fleshing out the parts, organization, and activities of a system, and how those give rise to behaviors of interest.  However, it remains somewhat controversial what it means for a model to describe a mechanism, and whether neural network models qualify as explanatory.

We argue that certain kinds of neural network models are actually good examples of mechanistic models, when the right notion of mechanistic mapping is deployed. Building on existing work on model-to-mechanism mapping (3M), we describe criteria delineating such a notion, which we call 3M++. These criteria require us, first, to identify a level of description that is both abstract but detailed enough to be ``runnable'', and then, to construct model-to-brain mappings using the same principles as those employed for brain-to-brain mapping across individuals.
Perhaps surprisingly, the abstractions required are those already in use in experimental neuroscience, and are of the kind deployed in the construction of more familiar computational models, just as the principles of inter-brain mappings are very much in the spirit of those already employed in the collection and analysis of data across animals.

In a companion paper, we address the relationship between optimization and intelligibility, in the context of functional evolutionary explanations. Taken together, mechanistic interpretations of computational models and the dependencies between form and function illuminated by optimization processes can help us to understand why brain systems are built they way they are.
\end{abstract}

\section{Mechanistic Explanations in (Neuro)science}
\label{sec:intro}

Due to recent progress in AI, deep neural networks (DNNs) can be trained to solve a variety of increasingly interesting tasks, including perceptual tasks that are natural for humans but historically difficult for artificial systems. A line of related work has proposed that DNNs can serve as quantitative cognitive models of behavioral patterns on such tasks, as well as computational neuroscience models of the brain systems underlying these human abilities.  But are such neural networks really \emph{explanatory models}, either of human and animal behavior, or neural activity in the brain?  Can we map them to the brain at all? Can they be used not just as tools for quantitative data analysis and curve fitting, but also to provide substantive scientific insights?

Defining what makes a model of a system as complex as the brain ``explanatory'' has itself been a challenging conceptual problem, whether for neural network models or any other kinds of models.  This problem is longstanding, but the recent quantitative successes of neural networks has raised the stakes.  If we can articulate a set of criteria to assess the explanatory value of models in neuroscience more generally, we can then apply what we find to a variety of examples -- most notably, the case of deep neural networks as models of the primate visual system.

Intuitively, an explanation should illuminate the \textit{dependencies} between the phenomenon of interest and the factors involved in generating it.\footnote{These dependencies and factors can be understood very broadly: everything from physical constituents to mathematical constraints, causal antecedents, or optimality considerations could in principle count as "factors upon which the phenomena of interest depend".  And of course there is no single goal for modeling in science; diverse goals support a diversity of explanatory uses. But we focus here on what we take to be a dominant mode of explanation in neuroscience, which involves a kind of functional decomposition of a system in order to explain how it produces behavior or capacities of interest.} Drawing on previous work in the philosophy of explanation, we start with the claim that to be ``explanatory'' for a target natural system, a model should:
\begin{enumerate}
    \item be a \textit{mechanistic} description of causal relationships in that target system, and 
    \item be \textit{intelligible} in the sense of being \textit{cognitively manipulable}
\end{enumerate}
Of course, it is somewhat controversial what it means for a model to describe a mechanism or be intelligible.  So in attempting to apply them in the complex context of the brain and modern computational methods, we will need to clarify our intended use of these terms, making clear the underlying theoretical and practical motivations. This first paper starts by dealing with the notion of \textit{mechanism}; its successor deals with the issue of \textit{intelligibility}.\footnote{See \url{https://arxiv.org/abs/2104.01489}.}

\begin{center}
***
\end{center}

\noindent In recent years, the vision of mechanistic explanation~\cite{machamer2000thinking, glennan1996mechanisms,Craverbook2007, levy2013abstraction} has been prominent in philosophy of science. This, together with the idea of interventionism~\cite{10.5555/331969, pittphilsci10974} as a means of gaining access to the causal structure of a system or phenomenon, is often presented as the goal of explanation in biology.

Broadly speaking, to construct a mechanistic explanation, we identify the relevant functional parts involved, articulate how those parts are organized, and how it is that their coordinated activities bring about the outcome or phenomenon of interest.  In the context of models in systems neuroscience, Kaplan and Craver (2011) have articulated this mechanistic ideal as a \emph{Model-Mechanism-Mapping} or ``\textbf{3M}'' constraint, to distinguish between those models that merely provide accurate descriptions or predictions of phenomena, and those that play genuine explanatory roles. 

According to 3M, a model of a target phenomenon explains that phenomenon when ``(a) the variables in the model correspond to identifiable components, activities, and organizational features of the target mechanism that produces, maintains, or underlies the phenomenon and (b) the (perhaps mathematical) dependencies posited among these (perhaps mathematical) variables in the model correspond to causal relations among the components of the target mechanism''\cite{kaplan2011explanatory}.

This account of mechanistic model-building captures a central endeavor of experimental systems neuroscience in practice, which attempts to explain capacities of interest by linking physiology to behavior.  However, it has been argued that computational models in neuroscience fit uneasily into the framework, if at all~\cite{ross2015dynamical, chirimuuta2017explanation}.  We believe that if enough weight is given to the central role of abstraction in the practice of such modeling in practice, a mapping criterion very much in the spirit of 3M \emph{can} serve as a unified framework for assessing computational models, including neural network models, as mechanisms for neuroscience.\footnote{Our discussion of models in perceptual neuroscience builds on the existing literature about the relationship between scientific abstraction and mechanistic modeling~\cite{levy2013abstraction, Piccinini2011, craver2018moredetails}.}

In particular, we note the intimate relationship between \textit{measurement} (and the choice of what to measure) in experimental neuroscience, on the one hand, and \textit{abstraction} (and the choice of what details to ignore) in computational neuroscience, on the other. We also note the importance of being able to compare data across animals, and survey extant strategies for doing so, all of which, again, require some details to be discarded and others to be preserved. In both cases, we make explicit the key role of abstraction that is already implicitly assumed in practice throughout systems neuroscience. 

Putting these two pieces together, we propose a version of the 3M requirement that \textit{highlights} the role that abstraction must play for quantitatively assessing the \textit{functional} similarity of a model to its explanatory target. 

To converge on the right level of abstraction, we propose that mechanistic models should be "runnable", in the sense of capturing enough of the causal structure of the phenomenon as to be able to reproduce them.  This helps to guarantee that in abstracting away from some details, we have nonetheless retained those features that are causally sufficient to generate the phenomenon of interest. 

Then, to capture the necessity of making comparisons across individuals, as well as mapping models to individual targets, we introduce the idea of a similarity transform.  Together, these ``3M++'' criteria (as we call them) are not only well-suited for application in assessing the value of the new generation of neural network models, but provide a unifying perspective from which to evaluate earlier (and now canonical) models from computational neuroscience as well. 

Here is a roadmap for the rest of this paper:  In Section 2 we discuss shared empirical foundations from neuroscience that underlie core abstractions that make the rest of thinking about neuroscience (and the ideas we talk about here) possible.  In Section 3, we discuss the original 3M criteria and make explicit two additional roles for abstraction in ``Predictively Adequate Runnable Abstraction'' and ``Transform Similarity'', both based in the foundational notions from neuroscience surveyed in Section 2.  Together, these broadened notions constitute what we call 3M++.  In Section 4, we develop a series of examples of 3M++, both positive and negative: models that illustrate either the presence or absence of mechanistic mappings. In Section 5, we return to our original motivating question: the case of DNNs.  We conclude that in some cases, DNNs can indeed provide reasonably good abstract mechanistic models, under the 3M++ criteria.  Finally, in Section 6, we draw some general lessons about finding the appropriate level(s) of abstraction for neuroscientific modeling.

\section{Foundational Abstractions from Neuroscience}
\label{sec:common_ground}

A caricature of two approaches to neuroscience might pit an industrious army of "more-details-better" experimentalists, painstakingly gathering as much data as possible, against the mounted horsemen of "do-more-with-less" computationalists, enchanted by the mathematical appeal of simple models for complex phenomena.  But of course, this \textit{is} a caricature, for (we claim) the two camps in fact share deep common foundations -- at least when the phenomenon they are both ultimately interested in explaining is the same.\footnote{Both camps would agree that different abstractions are likely needed to account for animal-level \textit{behavior}, for example, than to account for the biophysics of neurons \textit{qua} excitatory cells. The difference in outlook between the two camps might instead be captured as an expression of different kinds of humility: we don't know what matters, but if the brain exhibits it, it is probably important; we don't know what matters, but let's see how far we can get with minimal assumptions and a few variables.  And of course, given the tools available on either side, it may be that different aspects of the phenomena look like more promising candidates for the target of explanation.}

There are many cell types in the brain, and a plethora of biological activities. Every experimental neuroscientist is faced with choices about what to measure in the brain. What are the variables that are \textit{functionally} important in producing the behaviors or capacities of interest?  In practice, then, experimentalists are forced to make \emph{abstractions} that capture the details that \textit{matter} while ignoring those that don't.  It is this very activity that has led to a convergence on the central role of neurons (and neural activity) and their organization (including connectivity and functional localization).  

Of course, computationalists are also forced to deal in abstractions, perhaps more obviously so.  In order to check their models against data, they need to be able to make clear predictions, in order to make quantitative assessments of how accurate the models are.  While small models with relatively few components can be tested against data in a qualitative way, this becomes increasingly difficult as the number of components and interactions goes up.  At some point, it becomes necessary to have computational model that can be run on an external system rather than by hand -- and that requires precise formal descriptions of the relevant variables and interactions. 

But are the experimental and computational abstractions the same?  We claim that they often are -- and summarize this common ground below, juxtaposing the standard non-mathematical ``expermentalist'' descriptions with their formal mathematized ``computationalist'' counterparts.
Of course, it should be made clear at the outset that all of these abstractions are in a strong sense \textit{provisional}.  That is, they embody \textit{hypotheses} about what features of target system are important for explaining the phenomena.  If it turns out that they are inadequate, either because they neglect important features, so that we can not re-capture the phenomena of interest, or inaccurate, in a way that leads to poor predictions, then, like all hypotheses, they can and should be revised.  The adequacy and accuracy of these hypotheses requires empirical validation -- we have to see the abstractions in \textit{action} in a model that can be held responsible to empirical evidence, by making testable predictions about real brains.  

That said, although particular abstractions can and will be revised, together, they define an overall conceptual framework that is stable to individual changes in how we understand the precise role of particular components (or even what those components are).  This in turn implies that the general applicability of 3M++ for mapping such models to the brain is retained through changes to the particular abstractions employed.\footnote{... up to a point.  Of course, if we lose the idea that the brain is a network of cells whose spiking activities matter, we will also have lost most of the content of the framework.}  

\subsection{The Strong Neural Doctrine}
\label{subsec:snd}

The consensus in much of systems neuroscience is that neurons are what matter for behavior and in particular, the patterns of neural action potentials (or spikes) propagating through the brain, either intrinsic or evoked by external stimuli.\footnote{At least for behavior on short time-scales of a few hundred milliseconds to seconds.}  In this picture, neurons are functional units that respond to the spike rates of units synapsing onto them by aggregating those synaptic inputs (perhaps in a history-dependent way) and producing the appropriate spiking output by applying some relatively simple operation.

The strong neural doctrine, first inferred by Cajal from neuroanatomical observations~\cite{shepherd2015foundations}, is thus the idea that the brain is composed of a network of simple units. Now if all we care about are spikes and spike rates with respect to individual neurons, then all we care about are scalar-valued inputs and scalar-valued outputs (as opposed to vector-valued or multi-dimensional ones) to cells.  The \textit{propagation} of these spikes, in turn, depends on the idea that these cells are connected in a network. 
And the mathematical formalization of these ideas above describes the operation of a directed network of such units asynchronously applying their computation repeatedly, operating massively in parallel.  This is essentially the concept of parallel distributed processing (PDP) as introduced in the early 1940s by McCulloch and Pitts, and which also first gave artificial neural networks their name. 

More practically, what is often recorded and analyzed are spikes or their metabolic proxies, averaged in various ways: over varying time windows, across multiple cells, or over repeated trials. To the extent that a neuroscientist expects to explain their phenomenon of interest by appealing to such data, they are committed to the idea that this is an \textit{adequate} level of abstraction at which to describe their mechanism of interest, and moreover, that characterizing such spiking activity is sufficient to explain the behavior of interest. And indeed, much of the animal behavior that has been studied in a laboratory setting is well-predicted by these spike-rate averages.\footnote{The gap between those lab behaviors and complex behavior in the wild may still be significant.}

\subsection{The Canonical Neural Unit}
\label{subsec:ln}

The canonical description of a neuron is that of a functional unit that integrates its many synaptic inputs (excitatory or inhibitory, and of varying effective strengths), and then fires according to its biophysical properties whenever those inputs exceed some threshold. 
 
\begin{figure}
\centering
\includegraphics [width=.95\linewidth]{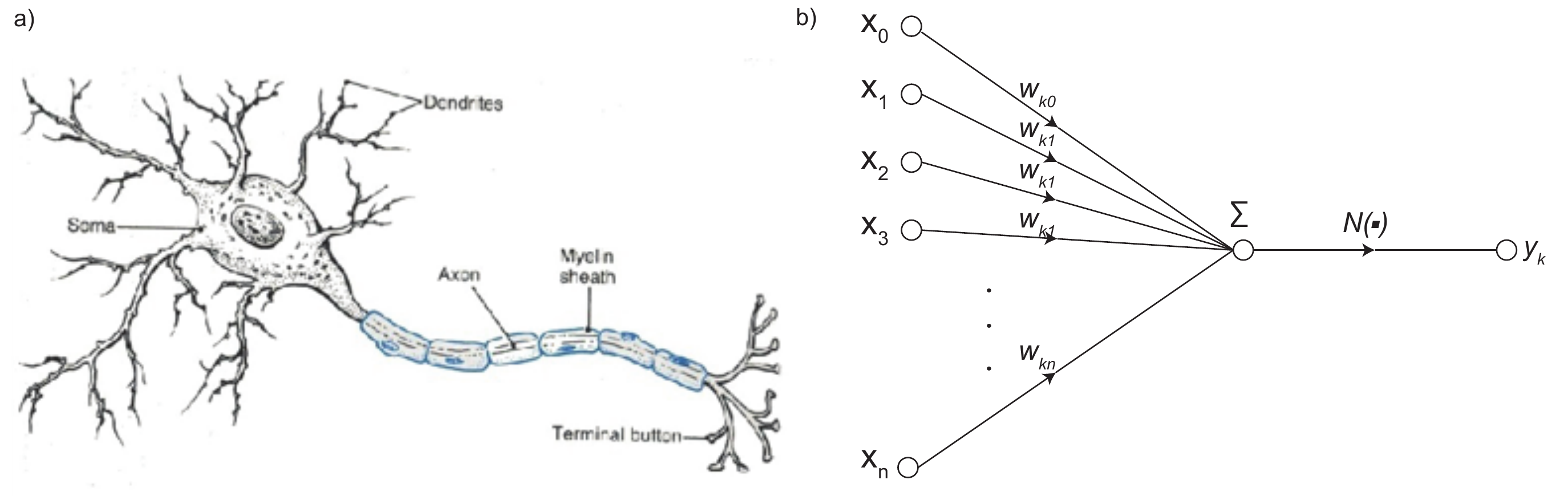}
\vspace{-3mm}
\caption{\textbf{a)} Standard dendrite-soma-axon picture of the neuron envisioned by neuroscientists. \textbf{b)} The Linear-Nonlinear unit supported by experimental data and used in standard neural network models.
\label{fig:ln_unit}}
\vspace{-3mm}
\end{figure}

A wide variety of experiments have suggested that, under normal physiological conditions, the neurons in the visual system can be described to some level of accuracy in a very simple form: that of the Linear-Nonlinear (LN) unit~\cite{pillow2008spatio,Carandini:2005kw}. The LN description is merely the formalization of experimental observations that purport to describe neural behavior at an accepted level of abstraction:  As a function of their neural (spike-rate) input, the neural (spike-rate) output of a neuron is well-predicted as a linear combination of its inputs (with to-be-determined weights), followed by some simple non-linearity, which in a biological neuron might, for example, manifest as a spiking threshold. (Fig. \ref{fig:ln_unit}).

Mathematically, this is equivalent to saying that:
\begin{equation} \label{eq:ln}
n_{i}(x) = N \left [\sum_{j \in \mathcal{S}_i} a_{ji}n_{j}(x) + b \right]
\end{equation}
where $n_{i}(x)$ represents the spike-rate output of a given neuron $i$ as a function of stimulus input $x$, $\mathcal{S}_i$ represents the set of all the neurons that synapse onto $i$, the numbers $a_{ij}$ and $b$ are constants, and $N[\cdot]$ is some fairly simple nonlinear function such as rectification. 

Let us re-emphasize that the LN formulation is a provisional abstraction: while it may be ``reasonably'' accurate, the LN form is definitely not a \emph{completely} accurate description of real neurons.  Nonetheless, three further points are worth re-emphasizing in the context of this particular example.

First, the LN description is a direct absorption by computational models of what brain scientists who are concerned with functional outputs have actually measured.  Since the specific version of neural networks typically used as neuroscience models are based on this LN approximation, to the extent that the LN form is inaccurate, neural network models of the brain based upon them will be as well -- but so too will be the many analyses in experimental neuroscience that presume that such a description is adequate for characterizing their data.

Second, if the inaccuracies of the LN approximation are sufficiently small or orthogonal to the capacity of interest, then models that posit a network of connected LN units could still be reasonably accurate for explaining that capacity of interest. 

Finally --- and perhaps most importantly --- if systems neurophysiologists of vision were to robustly validate a better approximation for the functional form of single neurons in the visual system (see e.g. \cite{poirazi2003pyramidal}), that new description could be substituted into all the arguments about using 3M++ to map models (such as DNNs) to the brain as described below, without substantially disturbing the form of the argument.

\subsection{Gross Functional Connectivity}
\label{subsec:cascade}

Throughout this paper, we use a part of the primate visual system known as the ventral visual pathway as our primary instance of an explanatory target. In addition to being one of the best-characterized complex multi-area brain systems in neuroscience, it is one for which the kind of explanation we argue for is especially apt.  

The ventral visual pathway solves a difficult ecological problem: that of rendering the ``blooming, buzzing confusion'' of incoming visual stimuli into internal states that subserve high-level behavioral goals such as scene understanding, navigation, and action planning~\cite{james1890principles}. We want to explain how this is achieved, in light of the patterns of neural activity observed in the system that we take to be involved.

\begin{figure}
\centering
\includegraphics [width=.95\linewidth]{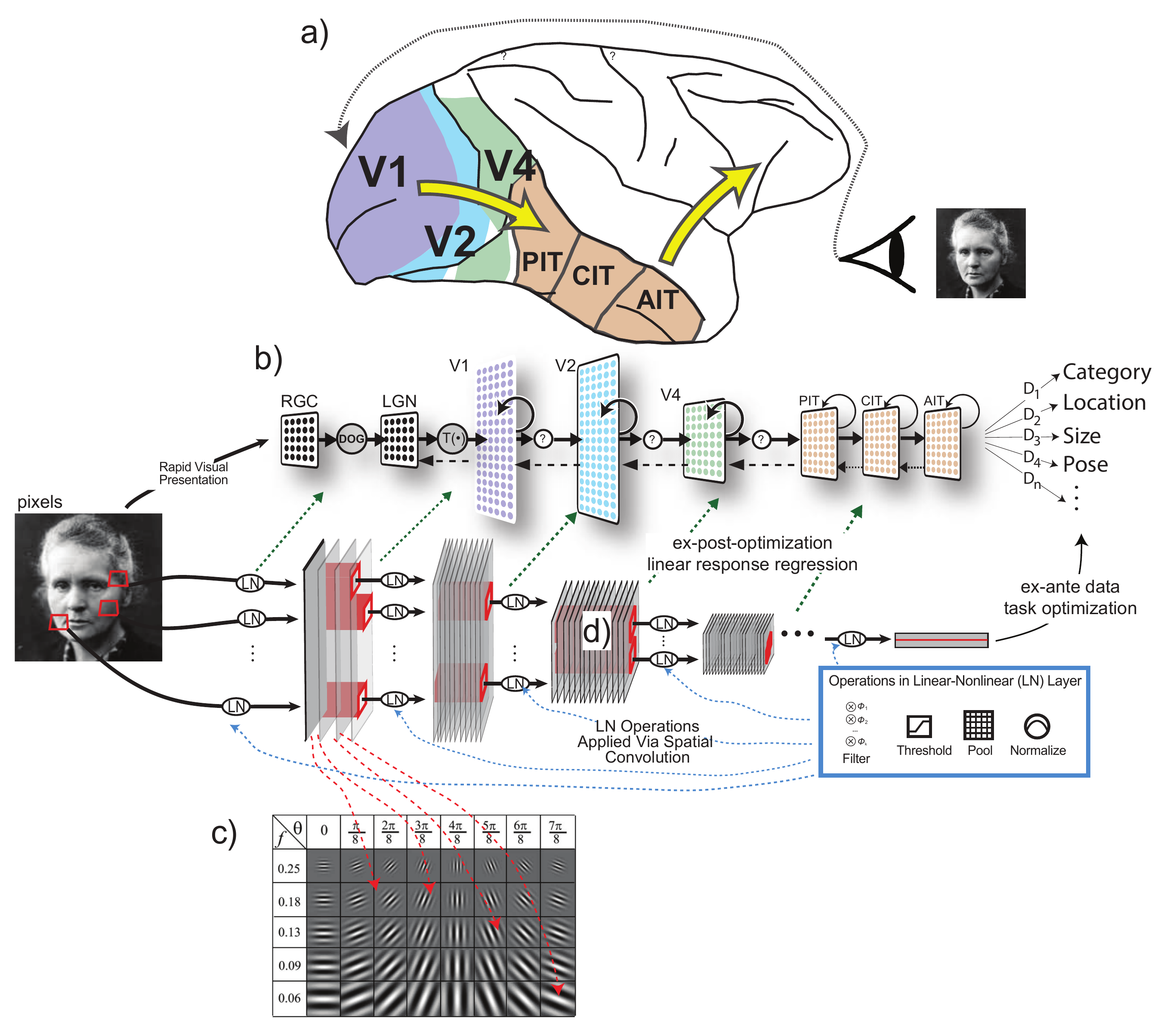}
\vspace{-3mm}
\caption{\textbf{The Primate Ventral Visual Pathway and Hierarchical Convolutional Models.}
\textbf{a.} The ventral visual pathway of humans and non-human primates consists of a series of connected cortical brain areas, roughly operating in a sensory cascade from early visual areas such as V1, through intermediate visual areas V2 and V4, to later visual areas such as inferior temporal (IT) cortex.  Neural responses in later ventral pathway areas are believed to encode a high-level representation of objects in visual stimuli, including infomation about category, position, pose, size, and so forth. \textbf{b.} Two key principles that have been extracted from neuroanatomical and neurophysiological study of the ventral pathway are that it is hierarchically and retinotopically organized. Hierarchicality means that retinal input activates a sequence of neural responses traveling downstream through the ventral pathway.  Retinotopy means that the neural activations in each area of the hierarchy are spatially distributed in correspondence with the input stimulus image, and described computationally by an identical (or at any rate, very similar) set of spatially local operations applied at each spatial location.  \textbf{c.} Gabor-filterbank convolution model of V1 cortex. The equation for any individual filter (the individual grey slices) is given by the Gabor wavelet equation, e.g. $g(x, y; f, \th) \sim exp \left [- \frac{x^2 + y^2}{2} \right] exp \left[ 2 \pi i f(x \cos \th + y \sin \th)\right]$, where $x,y$ are the coordinates in the input image and $f$ and $\th$ are, respectively, the frequency and orientation of the edge. The full filterbank contains frequencies $f$ and orientations $\th$ covering a range useful for extracting edges from natural images. \textbf{d.} A schematic of a hierarchical convolutional neural network and its conceptual mapping to the primate ventral visual pathway. 
\label{fig:ventral}}
\vspace{-3mm}
\end{figure}

In textbooks, the system is often presented as \textit{encoding} a stimulus in patterns of neural activity. That is, the retina transduces incoming stimulation into patterns of neural activity, which are then transformed as they propagate through successive later brain areas. \footnote{The visual system in humans and non-human primates is highly homologous – well-identified, anatomically distinguishable areas are primary visual cortex (V1), intermediate visual areas V2, V3 and V4, and inferior temporal cortex (IT). (See Fig. \ref{fig:ventral}a.) The cortical pathway beginning with V1 is connected to photoreceptor input in the eye via a series of subcortical pathways in the retina and lateral geniculate nucleus (LGN).  We will treat these subcortical circuits as an integral part of the system we seek to explain, but note the retina, LGN, and in fact the first cortical area V1, are not exclusive to the ventral pathway, but also subserve the dorsal pathway and other parts of the visual system.} 

Early visual areas, such as V1 cortex, respond well to low-level stimulus features including edges and center-surround patterns \cite{Carandini_2485, movshon1978spatial}.  Mid-level visual areas such as V2, V3, V4 and posterior IT (pIT) are less well-characterized by simple intuitive ascriptions than higher or lower visual areas closer to the sensorimotor periphery.  Nonetheless, neural activity in these intermediate areas appears to be at least in part describable as responses to features of an intermediate level of complexity between simple edges and complex objects, along a pipeline of increasing receptive field size~\cite{Freeman:2011gl, DiCarlo_2007, DiCarlo_2012, schmolesky_478, Lennie_2463, Schiller_49, gallant1996neural, Brincat_2008, yau2012curvature}.

The brain areas near the end of the ventral pathway can provide useful support for many different visual behaviors. That is, neural response patterns measured from IT cortex can be decoded --- by human scientists, and presumably by downstream brain areas themselves --- in the service of guiding a range of possible behaviors. Typically, IT is associated with robust object recognition~\cite{Hung:2005jh, majaj2015simple}. However, in addition to object category, attributes such as fine-grained within-category identification, object position, size, and pose, and complex lighting and material properties, can be easily decoded from neural activity in IT.\footnote{By ``easily decodable'' we specifically mean ``linearly decodable''.  The rationale for this is that proximate downstream neurons are only able to perform close-to-linear transformations on their input, and so any relevant features on which downstream neurons can be expected to condition their own outputs and behavior must be approximately linearly decodable from their inputs. (Not everyone agrees -- see for example ~\cite{rahnev_denison_2018}). Nonetheless, from the traditional neurophysiological perspective, ``easily decodable'' means that there is clear correlation that is relatively easy to extract without too much further processing, between features of interest and the response properties of the neurons. Of course, this usage of correlation essentially implies a linear readout model on the raw data --- as exemplified by the definition of ordinary least squares regression.}

While the processing stages in this cascade are simple, it is critical that they are at least somewhat nonlinear: the composition of linear operations is linear, so additional complexity can't be built up by a sequence of purely linear operations, and (plausibly) there would be no evolutionary point to allocating multiple brain areas for them in the first place. Given the non-linearities, complex transformations then arise from multiple such stages applied in series \cite{v4_fitting_reynolds_2012}. Since the original visual input is highly non-linear and tangled along the dimensions of the stimulus relevant for behavior, the untangling process by which the brain parses visual data is likely to be as well.

A successful model of the visual system then, should be able to reproduce its output, as well as accurately predict neural activities in these intermediate areas that are thought to subserve the production of that functionally relevant output.

\subsection{Spatial Locality and Retinotopy}
\label{subsec:conv}

Within the hierarchy of the ventral pathway, the activity in each distinct cortical area is itself retinotopically organized, so that the arrangement of neurons on the cortical surface mirrors the arrangement of their receptive fields (i.e. the parts of visual space that they are responsive to). In this way, neurons responsive to nearby regions of space are also located close to each other on the cortical surface. Together, the receptive fields of the neurons in each brain region completely cover the visual field,  and as we go up the hierarchy, receptive fields become larger (and thus less specific to a small region of visual space) as well as responsive to more complex visual properties in the environment. We also see that each part of visual space is treated more or less the same by the cells responsible for responding to it (at least within the foveal region) -- and this makes sense, given that any part of the scene may be in any cell's receptive field at a time, given eye and head movements.  See Fig. \ref{fig:ventral}b.

In other words, each area is separately organized as a spatially-distributed two-dimensional array, with activity in small-ish local regions within the array reflecting responses to inputs at a corresponding set of closely spatially-clustered locations in the original (two-dimensional) image stimulus. Each cortical area thus contains a spatial map of the whole visual field, composed of an array of these locally responding regions.The spatial resolution of this map decreases with each successive area within the visual pathway, from high resolution in V1 to low resolution in area IT.  Moreover, the distribution of response patterns as a function of input are observed to be very similar at each different spatial location within each area's spatial map.  Thus, each cortical area can be thought of as \emph{tiling} the overall visual field with a set of local response functions~\cite{hubel1962receptive, pasupathy2002population, gallant1996neural, dicarlo:tics_2007}.\footnote{We are only talking about foveal vision for the purposes of object recognition.}

In fact, the single most celebrated set of findings in visual neuroscience characterizes neurons in early cortical area V1 in essentially these terms. A line of experiments begun by Hubel and Wiesel~\cite{hubel1959receptive, hubel1962receptive} and carried further by many others~\cite{Carandini:2005kw, movshon1978spatial, ringach2002orientation} showed that V1 neurons can at least in part be described as responding optimally to ``edges'' of varying frequencies and orientations.

Over the course of several decades, it became clear that Hubel and Wiesel's array of `edge-detecting' V1 neurons could be redescribed in a very mathematically compact way: as performing spatial convolution with the so-called ``Gabor wavelet filterbank.'' Filterbank convolution is, by definition, the operation of multiplying each small patch of an input with a set of fixed patterns (the ``filters''), in a tiled fashion across an input array. This concept is a natural formalization of the spatially local, tiled, linear computation strongly suggested by the neurophysiological data. Models convolving input images with the Gabor filterbank, followed by simple rectification and normalization nonlinearities, have achieved striking success in characterizing V1 neural responses~\cite{Carandini:2005kw, ringach2002orientation, Carandini:2005kw, willmore2008berkeley}. See Fig. \ref{fig:ventral}c. While this characterization turns out to be quite a bit too simple given what we now know ~\cite{cadena2019deep}, it is nonetheless a good enough approximation to serve many useful predictive purposes. 

\subsection{Deep Hierarchical LN Cascades}
With all of the preceding abstractions on board, we can put the familiar pieces together to get something much less unfamiliar than it might initially have seemed. We've described the idea of a linear-nonlinear cascade, as well as the way that the linear contribution to neural response profiles within a region can be reformulated in mathematical terms as \emph{filterbank convolutions}. Taken together, the series of linear convolutions, interspersed with nonlinear operations, is a \emph{hierarchical convolutional neural network} (HCNN). See Fig. \ref{fig:ventral}d.

Though the filterbanks corresponding to early cortical areas were reasonably well-characterized by the Gabor wavelet model, it turned out to be quite difficult to identify a simple formula for filterbanks that would accurately describe activity in higher cortical areas. In fact, because of lack of success in extending the receptive-field-characterization program from early to higher cortical areas in the period up through (roughly) 2010, three key assumptions that were tacitly made in the above discussion began to be questioned. 

First, the hierarchical model might have been overly simplified.  Perhaps we also needed to take into account the (empirically observed) feedback and recurrent connectivity when describing neural responses in the visual pathway~\cite{gilbert2013top}. Second, perhaps the LN idea was too simple. Even if the relationship between inputs and neural responses~\cite{Carandini:2005kw} could often by characterized by an LN function and even if the relationship between neural responses and behavior could be expressed in the form of a linear decoder~\cite{majaj2015simple}, perhaps the functional relationships between neurons internal to the visual pathway were more complicated~\cite{poirazi2003pyramidal}. Finally, maybe the rate-code assumption was incorrect. Perhaps we had to return to a more detailed spike-timing description to adequately capture responses in higher cortical areas. 

At this point, it should be clear that these worries are essentially of the same form as earlier worries about whether we were employing the right abstractions. And just as before, they can be addressed by looking at the quality of the predictions they generate.  Success in characterizing the receptive field response properties of the visual system would help to allay these worries. As we will discuss in the next section, it turns out that deep hierarchical convolutional neural network models \emph{have} been able, in the last 10 years, to produce a substantial progress toward this end.

Before we get there however, we will return to the idea of model-mechanism-mapping (3M) and relate it to roles played in neuroscience by the shared abstractions just described.

\section{3M++: A Expanded Notion of Model-Mechanism Mapping}
\label{sec:mechanism}

\subsection{The Model-Mechanism-Mapping (3M) Constraint}

Recall that we are trying to employ the 3M requirement for evaluating explanatory models for neuroscience.  In its most straightforward form, it seems to give a clear formula for assessing whether a model describes a \textit{mechanism} - i.e. whether it is a mechanistic model. 

The relationship between mechanism and explanatory power is relatively intuitive, but perhaps worth making explicit.  When we identify the parts, activities, and organization of a system, we gain an intuitive understanding of its causal structure -- and moreover, a causal structure composed of concrete entities (in our case, biological) performing electromechanical activities familiar from other contexts, and for whose further explanation we may easily defer to well-established areas of science such as chemistry or physics.  There is a way in which mechanistic explanations are thus satisfying, in that they fit well into our existing conceptual structures. Happily, they are also the very explanations that most life sciences practitioners explicitly aim at.

That said, the \textit{exact} definition of what does or does not constitute a mechanism is somewhat contentious, precisely because of debate over the role of abstraction.  What exactly counts as an 'entity'?  Must it be spatiotemporally localized and contiguous?  How concrete must the characterization of its effects or activities be?  Sometimes, computational explanations are held out as prime examples of non-mechanistic explanation; other times, they are said to be merely incomplete mechanisms, sketches that must be filled in in order to qualify as full-fledged explanatory models.

To clarify these issues, we propose a  a revised version of 3M that gives an explicit place to common kinds of abstractions in neuroscience.  As presaged in the previous section, the kinds of abstractions commonly employed, far from being mechanistically disqualifying, are in fact the same ones relied upon by paradigm examples of mechanistic models.  That being so, with respect to this revised (3M++) condition, we think that some computational models are in fact prime examples of mechanistic explanation. 

\subsection{Predictively Adequate Runnable Abstraction}
\label{subsec:para}

Skeptics have taken it as obvious that no model-to-mechanism mapping can be found for NN models because the components, activities, and organizational features of those models seem very different from those of the brain -- ironically, given their name and historical origin. And it's true that the units in an artificial neural network abstract away from many apparently important features of neurons, such as their action potentials (spikes), and the fact that some are excitatory and others are inhibitory, while the standard gradient-descent-based techniques used to train artificial NNs are commonly considered to be biologically implausible.

But how important are these differences for our modeling purposes? Certainly we don't expect the same drugs to work on an artificial neural network as on a biological brain. Nor would we expect GPUs to respond the same way as brains do to changes in temperature or oxygen level. And if we zoom in far enough, no two animal brains are identical either. On the other hand, if we are primarily interested in a single clearly delineated capacity (say, immediate responses to visual stimuli under normal conditions), it might be that we can abstract away from these properties because these are not differences that make a difference in the context of that specific capacity, operationalized as performance on a particular task. There are many levels of description at which two systems may be similar, but one naturally illuminating and parsimonious level of abstraction is the \textit{least detailed} one that still captures the features functionally relevant to our capacity of interest, and is moreover adequate to predict the behavior of the target system which displays that capacity.

What we have stated explicitly here is already implicitly assumed in neuroscientific practice. To suppose that \textit{any} model system (even another biological one) could be similar to a target brain is to have already accepted that some details can be dropped and some differences of internal mechanism may be safely ignored. Which ones those are will depend on the phenomenon of interest.\footnote{Notice too that some degree of abstraction is already implicit in the notion of mechanism from which the 3M requirement arises.  Take the classical example of the mechanism of synaptic transmission. This mechanism has components at many different size scales: ion channels, vesicles, second messenger molecules, and so on.  The components of the mechanism are ``biological entities'' – not described at some maximal level of physical detail, but rather, those remaining after we have abstracted away from all physical details that are not functionally relevant to the biological phenomenon of interest, and where the explanatory target itself is characterized at some level of idealization.  (There is no synapse that looks exactly like the diagram in a textbook, and the predictions made by a textbook 3M-satisfying mechanistic model for synaptic transmission are for the most part qualitative rather than mathematically precise.)}

We have already emphasized that every experimental neuroscientist is faced with choices about what to measure in the brain.  To be undertaking neuroscience in a way that yields scientific insight requires us, like map-makers, to choose a level of abstraction far above that of the finest physical detail.   In section \ref{sec:common_ground}, we claim that modelers have for the most part made the \textit{same} choices as experimentalists in regard to these abstractions. But to what extent are these choices well-motivated in the first place?  In measuring (or modeling) brains, why shouldn't we have picked some other more or less detailed level of description than networks of spike-rate units? Or one that focused on different functional units? We claim that there is a reasonable answer to this question that imposes fairly strong constraints on what counts as a sufficient level of abstraction. 

The fundamental observation is that neuroscientists are interested in a particular \textit{task} or behavioral capacity, and the functional contribution of components of the system to the task of interest. For this reason, we want our model to be able to \textit{actually} perform that task. Thus, the level of abstract description we want must be one that still captures the features functionally relevant to our capacity of interest (and may vary depending on what that capacity is).  To this end, we introduce a notion of \underline{\textit{predictively adequate runnable abstraction}} (PARA), which requires that the model actually \textit{runs} successfully on novel instances of the same kind of input that the real system gets, to produce the output behavior of interest.  The requirement that the instances be novel rules out models that merely describe existing data, instead of capturing something important about how it is generated.

The capacity of interest itself may be characterized more or less finely – so that an abstraction that is predictively adequate to the capacity, understood coarsely, may no longer be adequate when we become interested in more details of performance (requiring us to re-operationalize the capacity with a finer-grained task). For example, if we are only interested maximizing the average accuracy across many ecologically-valid stimulus types, this places potentially less stringent constraints than accurately predicting the target's pattern of accuracy and errors for each stimulus individually. In practice, however, performance constraints can be very strong --- achieving high performance on ImageNet categorization imposes a high degree of consistency (although not complete consistency) with human error patterns. See for instance Rajalingham et al~\cite{rajalingham2018large}.

Critically, the requirement of runnability can end up forcing the preservation of many internal details.  That is, it can turn out that more details matter for the function of the system than we would have guessed from the armchair. If the model does run successfully, we may be reassured that we have not discarded more detail than we should have;  runnability validates something like the ``Salmon-completeness'' of the model being run. \footnote{ In Craver and Kaplan's terminology, ``the Salmon-complete constitutive mechanism for P versus P' is the set of all and only the factors constitutively relevant to P versus P'.''~\cite{craver2018moredetails}}  

Take the capacity of the visual system to categorize images by the objects they portray.  The criterion of runnability for the visual system model imposes strong constraints on how coarse the level of description can be.  It is very unlikely that a model that collapses all the distinction between multiple units within a cortical area and replaces them with a single ``area-level'' identity token would be sufficiently detailed to run at all. So the level of description that is typically used in analyses of high-level brain connectivity (e.g. \cite{bullmore2009complex}) may be informative about many scientific questions but is almost certainly insufficient for runnability on standard perceptual tasks.  A finer grain of detail -- at least preserving the ``unit-level'' of detail, with the capacity for having many variable units within each cortical area -- is needed for a model of visual behavior that runs based on pixel-level input. 

In a sense, then, a predictively adequate runnable abstraction is \textit{more} faithful than the kind of ``classical'' box-and-arrow mechanistic model often mentioned in the biological sciences, because we have imposed the further requirement that you must be able to run the model forward in order to predict – quantitatively – future states of the target system. This is especially important because it gives us a way to test the adequacy of the abstraction (as well as the causal relevance of the variables invoked), by directly assessing the accuracy of the output predictions.

We claim that such a predictively adequate abstraction is a useful way to elaborate on the 3M mechanistic requirement (call the new requirement 3M+). That is, if we use the notion of adequate abstraction to pick out just the functionally relevant components, activities, and organizational features, then a correspondence between these and components, activities, and features of the target system will satisfy the \textit{spirit} of 3M.

So far, our extension of 3M is quite modest. Many of the points regarding the ubiquity of abstraction have been explored by others~\cite{Stinson,Boone2016,craver2018moredetails,Piccinini2011}; we have just made explicit the kinds of neuroscientific abstraction that we will allow ourselves in constructing mechanistic NN models. In addition, we emphasize how important it is that the model be \textit{runnable} -- that it be not just a scientific representation of the capacity of interest, but a working instantiation of that capacity.

\subsection{Transform Similarity for Biological Populations}
\label{subsec:similarity}
With a notion of abstraction in hand, we turn now to the question of how to assess the extent to which a model ``corresponds to'' or resembles its target. Our starting point is that a baseline standard for resemblance is established by the similarity of one typical animal brain to another of the same species.\footnote{Similarity is a vexed notion. Everything is similar to everything else in some respects and not others.  Similarity fades in and out depending on our perspective, how closely we look, and which details we take to be important. See ~\cite{weisberg2012simulation} for a full discussion of the notion of similarity in the context of scientific modeling.} But this immediately raises a prior and subtler problem that must be addressed before we can really ask whether an artificial model is similar to the brain. This is the fact that there is no single primate visual system; every monkey and every human is different.

In fact, for most biological systems of interest, every individual of the species is different.\footnote{Even the \emph{same} individual can look different at different times.  Even in fully-developed adult brains, the exact roles of individual neurons can change over time. To give a very simple example, in the hippocampus, the well-known place cells that code an animal's location in its environment get \emph{remapped} regularly, and the same cell may not code for the same location before and after remappings.~\cite{ocko2018emergent} Some key regularities must remain constant over time -- place cells retain their identities as place cells over time, and distribution of locations coded for by the population of place cells remains the same -- or else nothing stable could be coded at all.  Nonetheless, the variability of the hippocampal population over time is substantial, and such variability is common throughout neuroscience. Thus, throughout the rest of this section, whenever we say ``different individual'', this encompasses the possibility of dealing with the same brain at different times.}  That means that to define a notion of similarity that is suitable for biological systems, we must deal with the fact that the target of explanation is inherently a \textit{population }of individuals. We must discover some sense in which the targets nonetheless share the "same" functional physiological organization, allowing us to map the components (and activities and organizational features) of one to those of another, even when they are not identical in fairly obvious ways. 

What allows us to do this is the acceptance of some kind of transform that supports the mapping from one animal to another –- and just as importantly for our purposes, the mapping from a neural network model to an animal.  It should be systematic, principled and not ad hoc.  Such transforms are already used in order to compare data across individuals (e.g. hyperalignment for fMRI data) -- we propose it as a means to assess model-target accuracy as well.  In order to come up with the right transform, the key question is: what transform class is needed to make accurate predictions about a given brain area of target animal $T$ on the basis of facts about the anatomically homologous brain area of source animal $S$? If $S$ and $T$ were exact duplicates of each other, we could predict the activity of every neuron in $T$ by observing neurons in $S$. But given that there is typically no one-to-one mapping of neurons from $S$ to $T$, we must open up the class of inter-animal transforms a bit wider.

In the case of the higher area of the visual system, such as V4 and IT, this transform class is often taken to be the set of linear maps.  That is, it is posited that the stimulus-driven activity of any neuron in a given area of individual $T$ can be reproduced by a linear combination of the activity of neurons in the analogous area of individual $S$.  Mathematically, this means that for all input stimuli $x$, the neuronal responses to $x$ by neuron $i$ in individual $T$, which we will denote $n_i^T(x)$, can be written as:
\begin{equation}
\label{eq:linear_transform}
n_i^T(x) = \sum_{j=1}^{M} a_{ji} n_j^S(x)
\end{equation}
where $j$ ranges over all $M$ neurons in the corresponding brain area of $S$. (This is the same as equation \ref{eq:ln} except there's no nonlinearity.) The numbers $a_{ji}$ are constants representing how much contribution each source neuron $n_j^S$ makes to replicating some target neuron $n_i^T$.  This is essentially a measure of similarity between the two neurons. If one source neuron made a hundred percent of the contribution to predicting the target neuron, then the mapping would be one-to-one. That multiple source neurons might need to contribute (i.e. $|a_{ji}| > 0$ for multiple $j$’s) is just a statement of the fact that there isn't a one-to-one mapping of the source animal’s neurons onto the target. 

But \emph{why} is this assumption about linear mapping made? The reasons are partially mathematical and partially empirical.  To understand the theoretical reason, we need to go back to a basic principle of the hierarchical LN cascade described in section \ref{subsec:cascade}.

While the processing stages in this cascade are simple, it is critical that they are at least somewhat nonlinear: the composition of linear operations is linear, so additional complexity can't be built up by a sequence of purely linear operations, and (plausibly) there would be no evolutionary point to allocating multiple brain areas for them in the first place. Given the non-linearities, complex transformations can only arise from multiple such stages applied in series \cite{v4_fitting_reynolds_2012}. Since the original visual input is highly non-linear and tangled along the dimensions of the stimulus relevant for behavior, the untangling process by which the brain parses visual data is likely to be as well.

To return to the problem of comparing two animals (assumed hereafter to be conspecifics and typical members of their population), we start with the uncontroversial assumption that they typically have the same number of constituent brain areas (i.e. processing stages) in their visual pathway (the linear-nonlinear cascade), and moreover, that the corresponding brain areas are themselves similar.  We can then ask, how do we assess the similarity of two corresponding stages in a Linear-Nonlinear cascade? Anticipating the application of this question to neural network models later, we'll call these stages \textit{layers} - with the caveat that these are meant to suggest the layers in a NN, and do not refer to the cortical layers anatomically arranged \textit{within} a given brain area.

We can make the comparison simpler by noting that the nonlinearities are drawn from a limited number of simple (unparameterized) forms corresponding to discrete functional ``cell types'' that are plausibly the same for any two animals. Now our question has been reduced to the question of when the linear portion of the transform is the same.

The simplest and strictest notion of similarity across corresponding brain areas in two animals would require that the two sets of neurons can be placed in one-to-one correspondence, i.e. that there is a permutation of the indices of one animal's neurons such that it becomes equal to the other's. 
In early areas along the visual pathway, such as the retina, this strict mapping may largely be correct.   
But for intermediate and higher cortical areas such as V4 and IT, where neurons will be highly experience-dependent and represent objects and parts of objects, such a one-to-one mapping is almost certainly too strict. 

Instead, a natural next-strictest step is to say that sets of neurons in such areas are similar when one can be constructed from the other via an invertible linear transform.  When this is possible, we will have used the linear transform as a remapping between the two animals. Or equivalently, we've taken one animal to be a ``linear model'' for the other.

The class of linear transforms is more powerful than just permutations, and thus defines a looser equivalence notion than one-to-one mappability.  But the linear transform class is not all-powerful; for example, it is \textit{not} the case that just \textit{any} two groups of neurons can be transformed to each other via linear transform. In fact, linear transforms are comparatively weak, because (and this is obvious once you think about it) any two brain areas that are related to each other by a \textit{non-linear} relationship (as we believe V4 and IT to be) are by definition \textit{not} linearly transformable to each other.  

A convenient consequence of this ``weakness'' of linear transforms (and thus the relative strictness of the similarity class they define) is that performance on any task decoded by the modeler from linearly equivalent areas will be the same, assuming that a simple linear decoder is used to assess performance.  For example, if one uses a standard method such as a support vector machine (SVM) or linear discriminant analysis (LDA) to assess the explicitly accessible information for a given task that is present in a given brain area, then performing a linear transform on the data will not change the results. Thus, if two individuals are said to be equivalent when their neural responses in corresponding areas line up under linear transform, this means that the individual’s performance on any behavioral task (assuming the brain areas involved are actually being used to support that task behaviorally) will be the same. The linear transform class is essentially the largest (and thus least assumption-prone) transform class that has this property.

Aside from these mathematical reasons, the choice is also somewhat empirically-motivated.  
To be clear, it is not yet known empirically whether response properties in two animals’ brain areas really are equivalent up to linear transform.  The reason this is not known is simply that no single experiment has yet amassed enough neural data in a single cleanly-comparable experimental condition to ascertain this fact, though it has not yet been ruled out.  But what is known empirically~\cite{Kriegeskorte_2526,majaj2015simple} is that activity patterns in corresponding brain areas between animals are substantially more similar to each other up to linear transform than are the activity patterns in two different brain areas within the same animal.  That is, for brain areas V4 and IT and source/target animals $S$ and $T$:
$$dist(\text{V4}^S, \text{V4}^T) \text{ and } dist(\text{IT}^S, \text{IT}^T) \ll dist(\text{V4}^S, \text{IT}^S) \text{ and } dist(\text{V4}^T, \text{IT}^T)$$
where $dist(\cdot, \cdot)$ means distance in linear span.  Similar empirical inequalities have been shown to hold when comparing other pairs of areas in the ventral pathway, e.g. V1 vs V4 or V1 vs IT. In fact, the distances between pairs of areas mirrors the structure of the ventral pathway itself, with $dist(\text{V1}, \text{V4}) , dist(\text{V4}, \text{IT}) < dist(\text{V1}, \text{IT})$. 

Linear transform thus does appear to empirically capture a reasonably effective notion of similarity in which the abstract structure that is supposed to be the target of description (``V4'' or ``IT'' in ``the brain'') is well defined, up to linear transform. In mathematical terms, we can ``quotient out'' the inter-animal differences by means of the proper class of inter-animal transforms, creating a ``population equivalence class'' as the target of explanation.

While linear transforms are a reasonable first pass, we are \textit{not} claiming that linear transform is the best or only concept of similarity in any deep sense --- only that \textit{whatever} turns out to be the strictest (e.g. least expressive) empirical mapping class demanded by the population is what should be used to define similarity between different individuals.  Linear mappings are just the current simplest transform class that appears to be consistent with the known data, and the pseudo-mathematical argument above is just a heuristic that explains why this might be true.

However, were a very large set of ventral stream neural responses on a common set of images to be collected from a sufficiently numerous population of macaques, it would become empirically feasible to determine the actual inter-animal mapping class.  If it turned out that the full strength of linear maps was needed to map one animal to the other, the choices described above would be justified. However, if animals were more similar than linear (e.g. perhaps only orthonormal transforms are required), or animals were less similar than linear (e.g. non-linear mappings are required), the choice of linearity would have to be revisited, and the specific results in the literature that depend on this choice would also need to be revised -- but the overall structure of our argument here would remain (much as the argument in \S\ref{subsec:para} would remain even if neuroscientists validated a better model for neurons than the LN abstraction).\footnote{In fact, the empirical data from various areas in the visual pathway give clues to ways in which stricter, more "one-to-one"-like, classes of transforms are probably indicated.  As mentioned above, in the retina, at the very beginning of the visual pathway, neuronal circuits appear to be highly stereotyped within species, and it is perhaps possible that, for any retinal neuron in one animal, a single neuron could be found in any other animal of the same species that matches the original neuron's response pattern very closely. Somewhat further away from the sensory periphery in cortical area V1, a strict one-to-one match between organisms would be much less effective (especially since different individuals can have different sized V1s~\cite{schwarzkopf2011surface}). Nonetheless, the  strictest transform class between different animals' V1s must be in some sense narrower than the full space of linear transforms, because there is no guarantee that the stereotypical V1 simple and complex cells that are so robustly observed would be preserved under linear transform (since a linear transform of a Gabor wavelet filterbank needn't contain any Gabor wavelets).  Going further down the ventral pathway, to (e.g.) V4 or IT, the flexibility of the linear transform class is likely to be increasingly important -- the further one gets from the sensory periphery, as layers of non-linear synapses pile up, the more slip-room there is likely to be in how different the response patterns are between individuals.  However, even in IT cortex far from the sensory periphery, the existence of large fractions of units with particular selectivity patterns (e.g. the neurons in the well-known face, body, and place areas~\cite{Downing:2001dr,Kanwisher_546, Epstein:1998uj}) shows that the correct transform class can't \emph{quite} be linear.  That is because the fraction of units in a population with a particular selectivity is not a linear invariant: a linear transform can too-loosely up- or down-weight the prevalence of units with specific selectivities, changing the selectivity profile of the population in abiological ways.  

However, despite all these ways in which the linear transform class might be too loose to properly account for the inter-animal similarities in various visual areas, it is still a much better mapping class than the obvious alternative -- namely, the much stricter one-to-one mapping. Hopefully as the field progresses, understandings of the proper transform classes for each visual area better than either one-to-one or full-linear will emerge.} 

So far in this section, we have been concerned with defining the scientific target phenomenon itself. But now returning to the question of comparing models to the brain, we need to modify the definition of a mechanistic model given by the 3M constraint to accommodate the nature of the target.
We propose that \textit{whatever class of transforms is used to define similarity between individuals should also be the one used to make mechanistic mappings between mathematical models of the system and any one target individual in the population}.  Thus if we decide based on empirical grounds that the proper class of transforms between individuals in the population is linear, then the proper notion of a mechanistic mapping between a model and any one individual should also then be linear.  In other words, we should be able to write the responses of any monkey neuron in terms of a linear combination of units in the model,
\begin{equation}
\label{eq:synth}
n_i^T(x) \sim n_{\text{synth}}(x) = \sum_{j=1}^{M_{\text{model}}} a_{ji} n_j^{\text{model}}(x)
\end{equation}
where $M_{\text{model}}$ is the number of units in the model layer.  We can think of $n_{\text{synth}}$ as a ``synthetic neuron'' (created from the linear combination of model units) that maps directly (one-to-one) to a monkey neuron.

It might be tempting to object that the modeling of target real neuron $n_i^T$ as an amalgam of artificial neurons in the model doesn’t \textit{feel} very ``mechanistic''.  But remember that this same formula is the very one needed to even determine the sense in which a target neuron is similar to neurons in the same brain area of another animal of the same species. To the extent that activity in brain areas across individuals are predictable phenomena in the first place, the modeling relation expressed by eq. \ref{eq:synth} is as mechanistic as it can ever get (unless our aim is to build a \textit{different} model for each individual).\footnote{A consequence of this idea is that, instead of requiring that a computational model perfectly predict activity in a target system, we should only judge accuracy up to the level at which one population member (one animal) can predict another population member (a conspecific). This defines a clear ``noise ceiling'' with which to normalize prediction accuracy: even if the literal match between predictions of a model and a given individual is imperfect (i.e. a correlation of less than 1.0), as long as this similarity is as good as the typical individual-individual similarity (which itself might exhibit intra-specific correlations of less than 1.0), the model can be said to be predictively adequate.}

Adding this second elaboration to what kinds of abstraction make sense in constructing our model-to-mechanism mapping, we arrive at a set of revised criteria we'll call "3M++".  And just as the requirement for runnability discussed in the last section imposes some constraints on the level of detail of internal structure that can allowably be left out of an acceptable model, so too does the requirement that the model-to-target mapping be of the same kind that we use to map individuals to each other impose some sharp constraints on acceptable models.

To illustrate the kind of model that would be ruled out, consider the famous result in theoretical neuroscience called the Universal Approximation Theorem (UAT)~\cite{cybenko1989approximation}. The UAT (roughly) says that any functionality that can be generated by a deep multi-layer neural network can be approximately by a shallow network with a single, potentially very large, hidden layer. However, even if such a single-layer network passed the functional adequacy criterion of \S\ref{subsec:para}, it would immediately fail the mappability criterion discussed here, since the kind of mapping required to go from a huge shallow network to a multi-layer network with fewer units doing the same task is \textit{not} the same kind of mapping that would be needed to use data from one monkey to predict activity in another. In fact, recent mathematical analyses suggest that the number of units required by a shallow network to replicate the functionality of a deep network is exponentially greater -- e.g. the complexity (in a suitable sense) of the functions computable by ANNs grows linearly with network width (number of units per layer) but exponentially with depth (number of layers).  Thus, the violation of the transform similarity constraint would be very large. See \cite{poggio2017and} for a discussion of these ideas.

\subsection{The 3M++ Criteria}

\begin{figure}
\centering
\includegraphics [width=.95\linewidth]{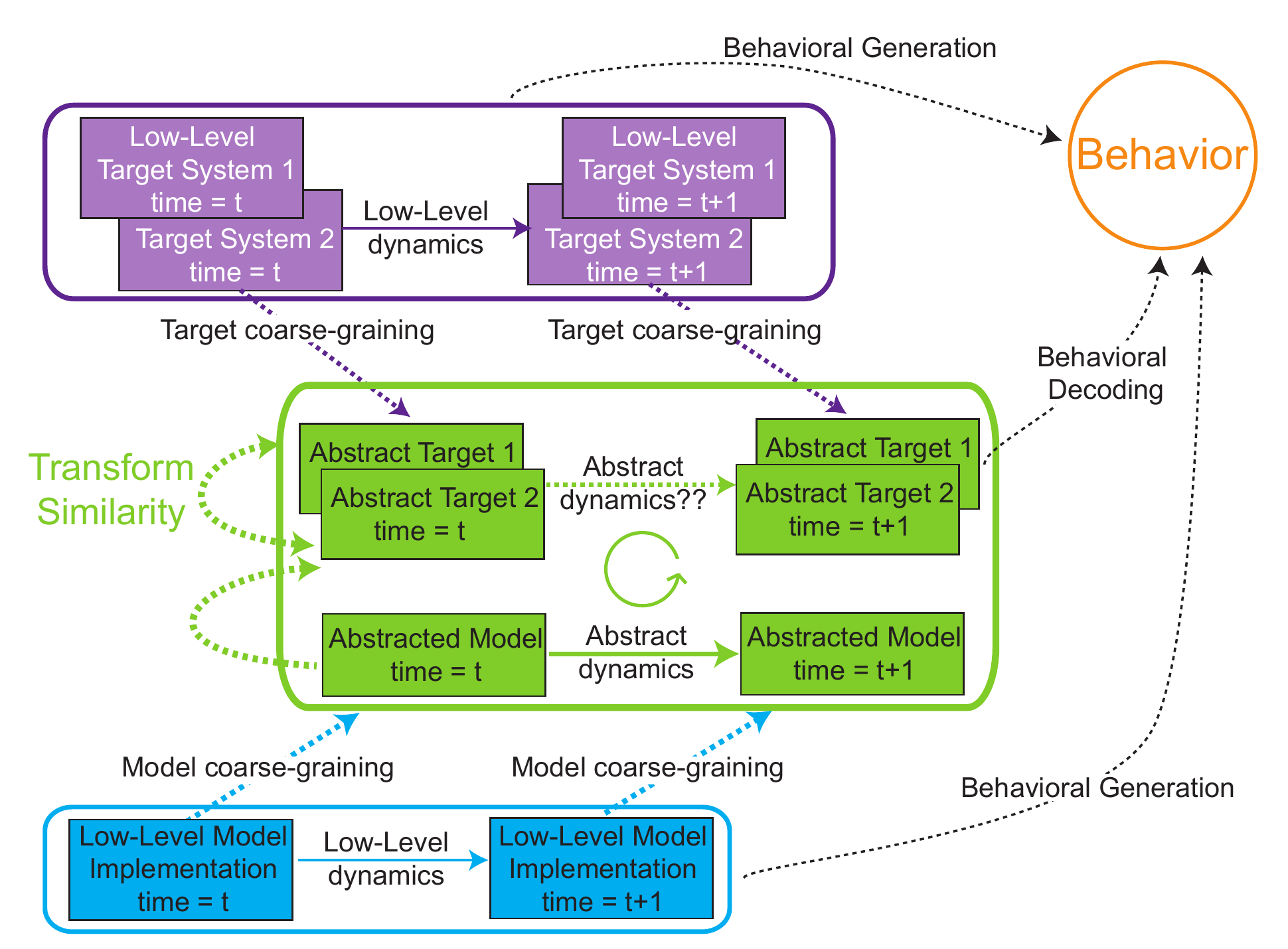}
\caption{
\footnotesize \textbf{The 3M++ Concept.} The overall goal is to  mechanistically describe and explain a target physical system (purple box), at whatever level of detail that is just sufficiently fine-grained to account for some chosen behavioral output (orange circle) generated by the target system. The target system itself is not really one system, but instead a population of individuals that are somewhat organizationally distinct (exemplified by Target System 1 and Target System 2). The model used to explain the target system (blue box) is a computationally-precise fully-runnable implementation that itself also generates the behavior of interest. The mapping between the model and the target is mediated by a common abstraction (green box) to which each is mapped via type-specific, information-losing and thus non-invertible coarse-grainings (purple and blue dotted arrows). On the one hand, the abstraction must be coarse enough that it can put systems with two very different low-level implementations into registration in the same state space. On the other hand, the abstraction must be fine-grained enough that it can actually support the behavior of interest under some reasonable simple decoding scheme. Moreover, passing to the abstraction is necessary to capture the sense in which the target system is well-defined, because the information-preserving, largely invertible similarity transforms needed to map instances of the target to each other (dotted green double-headed arrow) are only defined in this abstract space. (And this too places constraints on the abstraction level.) The relationship between the abstracted model and the abstracted target system (represented by the single-headed dotted green arrow) is then required to be of the same type as the transform similarity needed to map (abstract) target instances to each other. To the extent that the (abstract) dynamics of the abstract model correctly describe the changes in (abstract) state of the target system, the model is predictively adequate --- that is, the green portion of the diagram ``commutes'' (as represented by the circular green arrow). When all of these conditions hold, the 3M++ criteria are met. A model that satisfies the criteria  provides a proof-by-example that the dynamics of the target system can be captured abstractly (i.e. that the mapping represented by the arrow labeled ``Abstract dynamics??'' is well-defined). Prior to having a 3M++-satisfying description in hand, it may be difficult or impossible to directly observe the low-level dynamics of the real system sufficiently completely to assess at what level it supports abstraction the first place.  
\label{fig:abstraction}}
\end{figure}

The force of the ``transform similarity'' expansion of 3M described in \S\ref{subsec:similarity} is to allow for a class of systems with the same types of underlying components (e.g. neurons) to be considered a single explanatory target, even when any two observable instances of the system are different in some substantial regard (e.g. having responses that differ up to a linear transform). The force of the ``predictively adequate runnable abstraction'' expansion of 3M (\S\ref{subsec:para}) is to allow for systems with different very-low-level implementations (GPU transistors vs biological cells) to be compared at a meaningful functional level.

This distinction can be visualized as a diagram (Fig. \ref{fig:abstraction}) in which the abstraction mappings are the dotted diagonal arrows that attempt to bring implementations with different physical details (blue vs. purple) into registration at a shared level of description (green), while the similarity mappings are the green curved dotted arrows that attempt to bring different abstract descriptions (either of individuals within a biological population, or of a model and a target system) into registration. The abstraction mappings are non-invertible, and differ from implementation type to implementation type: you lose detail when you go from GPUs to activations in the abstract description and these are \emph{different types} of details than the ones you lose when you go from biological cells to activations. 

Transform similarity by contrast is invertible and should be the same whether the sources are models or biological instances: insofar as you are required to discard details in going from data collected from Animal $S$ to data collected from Animal $T$, the same exact sort of discardings must be demanded when going from  Animal $T$ to Animal $S$, or from abstract descriptions of model $M$ to abstract descriptions of any animal instance, or vice-versa.

Moreover, the constraint imposed by the mapping similarity requirement interacts with the constraints imposed by the predictive adequacy criterion, which require that the model be articulated and detailed enough that it can be run forward to predict the activity of the target system, on the same kinds of stimuli as those to which we want to explain the target system's responses.  This comes to a way of helping us locate the right level of abstraction to explain the phenomenon of interest. First, in describing what we are interested in explaining, we fix the level of grain of the \textit{explanandum} in part by loosening the description of that explanandum until it can capture what is \textit{shared} among members of a population (e.g. "what can be mapped by such-and-such similarity transform to a shared space"). Then, we want to capture all the details needed to explain \textit{that} explanandum -- but no more.  (Just as when building a minimal model, we want to throw away as much detail as possible -- but no more). Runnability serves to \textit{validate} the abstractions chosen, guaranteeing that they are \textit{sufficient} to reproduce the phenomenon of interest, at the level of generality desired.

To illustrate using the same examples given earlier:  the PARA constraint would rule out very coarse models that collapse the need for multiple units within each area as potential mechanistic models for the monkey visual system. 
Meanwhile, the similarity mapping constraint would rule out a huge shallow network, which would keep the units intact but collapse the organization structure of brain areas. 
Together, the two impose significant constraints on how abstract or intuitively unlike the brain models can really be -- indeed, likely stronger constraints than those articulated in the original 3M criteria for what constitutes a mechanistic model.

\section{Some Illustrative Examples of 3M++}
We turn now to some cases to illustrate the contrast between the presence and absence of genuinely deep similarity between a model and the phenomena it predicts, to flesh out how the 3M++ criteria work in practice. 

\textbf{Neural Networks in Genomics (Negative example):} Recent work in computational genomics has shown that neural networks can be successfully fit to various aspects of genotype-to-phenotype relationships.  For example, the state-of-the-art prediction of 3D protein folding structure from the protein's 1D genetic sequence is currently done by deep neural networks~\cite{wei2019protein}.  In this case, there is no claim that the internal layers of the neural network have any kind of part-level mapping to the biological phenomena that actually give rise to the protein (e.g. either the ribosome that mechanistically builds the protein on the basis of the genetic sequence or the energy-landscape considerations that cause proteins to have reliable folding dynamics).  These were never intended to be mechanistic models, and the level of coarse-graining at which a ``part-level mapping'' would be possible would completely quotient out all the details of the network into a single ``box'', at which level the model would no longer be runnable. Thus, the 3M++ criterion fails in this case.

\textbf{Single Neurons (Positive):} Returning to neuroscience, it is instructive to first look at the case of a single electrically-isolated neuron. 

To illustrate a positive example of the application of our criteria to a well-known model, we'll describe a disagreement over whether the Hodgkin-Huxley model of neural firing qualifies as a mechanistic model.  Hodgkin and Huxley famously proposed their model of neural firing on the basis of the interaction of several ionic currents by fitting exponential functions to observed voltage changes in giant squid axons. Their model matched the data (having been constructed from it), but also had provocative features that suggested as-yet-undiscovered features of the structure of the sodium and potassium-specific ion channels.

One of the main exponents of the mechanistic framework (Craver, 2006) has argued the Hodgkin-Huxley model is not a full mechanistic model, because it appeals to undischarged filler-terms such as "activation" or "inactivation" (particles).  In response, Levy (2013) points to the role of deliberate abstraction for the purposes of describing a phenomenon at the relevant level, as well as the ubiquity of ``aggregative abstraction'' in the description of biological phenomena at all scales.  So for example, when tracking phenomena in population genetics, or trying to explain cellular physiology, we must often consider not single concrete entities as the relevant "entities" in our mechanism, but rather more generalized "things" like ionic fluxes, the "spread" of an allele, or averages and distributions of smaller distributed entities.  We agree with Levy that the abstractions indispensably employed throughout systems neuroscience have this distributed, partially mathematized flavor. 
Nonetheless, all parties agree that \textit{once} we agree to interpret the relevant variables as mapping to subunits of the ion channels (at least provisionally, in a way that, again, can be revised if needed), then the resulting model is clearly mechanistic.\cite{craver2018moredetails} And in the decades since Hodgkin and Huxley, the details of models of neural firing have been developed substantially, resulting in detailed \textit{compartment models} that confirm and build upon the initial mechanistic mapping.

One cell type that has been the target of detailed modeling work is the cortical layer 5 pyramidal cell (L5PC).  Painstaking low-level electrophysiological measurements have shown how the electrical responses of L5PCs are well-characterized by a biophysical multi-compartment model~\cite{hay2011models} that elaborates the original Hodgkin-Huxley model, but expands the list of specific parts in the model, and the accuracy with which they are mapped.  For the same reasons that the HH model meets the 3M++ criteria once its components are mapped, so too do these multi-compartment models.  In fact, such biophysical models are sufficiently accurate, at least in isolated circumstances, that they used to serve as ``ground-truth'' descriptions of L5PCs for other modeling purposes, as we'll describe next.

\textbf{Single Neurons (Negative):} In recent work~\cite{david2019single} David et al. construct a deep neural network in order to produce a fine-grained model of a \textit{single neuron}.  Specifically, the authors show that it is possible to use deep neural networks with fully connected layers -- also known as multi-layer perceptrons (MLPs) -- to reproduce the input-output profile of (the multi-compartment model of) L5PCs.  The MLP models are then created by fitting the weight parameters to the responses generated by this ``ground-truth'' biophysical compartment model. The authors find that a seven-layer MLP can learn this I/O relationship effectively, achieving high performance on held-out testing data.

Though it is predictively adequate, the seven-layer MLP does \emph{not} satisfy the 3M++ criterion because it does not satisfy a model-to-mechanism mapping.  This is because the components of the MLP do not map to real-world components. No physical or explanatory significance can be assigned to the weight parameters of the MLP. The seven fully-collected layers in the model do not correspond to seven physical cellular subsystems connected in series within the (ground-truth) L5PC cell biophysical compartment model. Nor do the the components of each layer (the artificial neurons) correspond to physical entities such as molecules or molecular aggregations that are components of the biological cell being modeled. Indeed, the finest coarse-graining at which it would be possible to collapse the 7-layer MLP model and the L5PC compartment model so that there would be a correspondence would be one where we had abstracted away \emph{all} these non-corresponding details with a very coarse coarse-graining --- at which point neither model would be runnable.

\textbf{Gabor Model of V1 (Positive):}  The observation that the Gabor filterbank model of V1 (Fig. \ref{fig:ventral}c) is an abstract mechanistic model according to 3M++ largely boils down to restating the results described in section \ref{subsec:conv}.  It is evidence that the Gabor model is runnable, since there is a well-defined algorithm for computing output responses for any arbitrary input image.  Furthermore, to the extent that Gabor wavelets actually do a reasonable job at predicting real V1 neural responses (i.e. well, but not perfectly), the model is somewhat predictively accurate.  The Gabor model can be written in the form of a neural network with one hidden layer, and the mapping from the scalar-valued activities of the artificial neurons to the spike-rates of real V1 neurons is direct.  In fact, unlike in higher cortical areas such as V4 and IT, where a looser transform class (e.g. linear transforms) is required to identify similarity between neural samples, in V1 models (and V1 data), it is likely that a single neuron-to-neuron mapping is possible -- at least, for those V1 neurons that the Gabor adequately predicts in the first place. 

\section{The Case of Deep Neural Networks}
\label{subsec:nns_are_para}

Recall that the original 3M criteria required us to be able to map model components to brain components, model organization to brain organization, and model activities to brain activities.  If we understand this as the model satisfying 3M++, so that both it and the target system can be mapped to a common class of runnable abstractions, we think that deep HCNN models of the ventral visual pathway can, in certain circumstances, count as mechanistic models.  
To see this, we'll check the requirements needed to establish 3M++ as exemplified in fig. \ref{fig:abstraction}.

On the one hand, starting with the model (the blue boxes in fig. \ref{fig:abstraction}), the HCNN is constructed to be runnable: it is a well-defined computational program accepting as input any image-like stimulus, and performing the ecologically-relevant task that is a proxy for the capacity of interest. Second, there is (obviously) a coarse-graining from the components of the physically-implemented HCNN running on (say) a GPU to a more generic abstract level of description. In other words, we've passed from the blue boxes in Fig. \ref{fig:abstraction} to the bottom row of green boxes (the ``abstracted model'') in that figure. 
   
Turning to the target biological system (the purple boxes in fig. \ref{fig:abstraction}), the arguments from section \ref{sec:common_ground} can be rephrased as saying that there is a coarse-graining from the components of any instance of a real brain's ventral visual system to that \textit{same} abstract level of description as used for the model.  
Once at that shared level of abstraction, we can use the same transform as that used to register the activity of any two animal brains to map and predict the activity of components in the model to the activity of components in the real brain, in a way that respects the organizational structure of both the HCNN and the ventral pathway.  In other words, we've constructed the top row of green boxes (the ``abstracted target'') in Fig \ref{fig:abstraction}, and have posited that the dotted green arrows exist. 

But there is one major ingredient that is missing: we have not shown that the abstract dynamics of the two systems (e.g. the target and model dynamics) are actually similar.\footnote{In mathematical terms, does the inner diagram actually commute?}.   
To address this, we arrive finally at the place where the recent results from comparing modern HCNNs to neurophysiology are brought to bear. Indeed, everything up to this point could actually largely have been said about the very earliest multi-layer convolutional networks, such as the neocognitron~\cite{fukushima1980neocognitron}. At this point, however, we need to ask whether the \textit{activities} in the layers of the network correspond to neural activities in brain areas throughout the ventral pathway -- and that is where modern HCNNs have succeeded to an extent that earlier models did not. 

In these results, a deep HCNN is built with about the the right number of layers as there are observed brain areas in the ventral visual pathway. Then the parameters of the HCNN are optimized such that the resulting network is able to solve a challenging behavioral task -- typically, 1000-way object categorization in real-world images~\cite{Deng_3067}. Intriguingly, it turns out that the \emph{error patterns} of the outputs of these goal-optimized networks correspond to a large degree to those measured from data in human and primate behavioral experiments~\cite{rajalingham2018large}, even though the networks were optimized for overall performance rather than for producing any particular error pattern.  Most importantly for our purposes, it also turns out that HCNNs built this way are by a very long margin the best quantitative models of neural responses in every measured cortical area of the primate ventral visual pathway. Specifically:

\begin{figure}
\centering
\includegraphics [width=\linewidth]{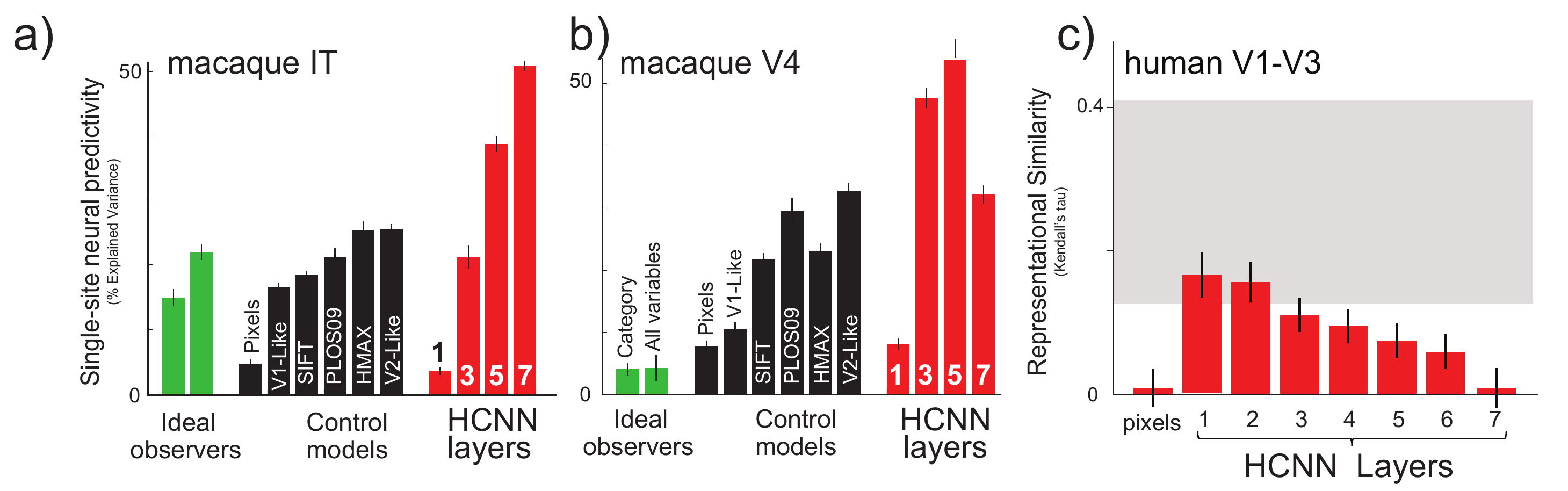}
\caption{\textbf{Quantitative comparison of HCNN model with primate visual areas} \textbf{(a.)} Based on~\cite{yamins:pnas2014}, comparison of ability of various computational models to predict neural responses of populations of macaque IT neurons (right). The HCNN model (red bars) is a significant improvement in neural response prediction compared to previous models (black bars) and task ideal-observers (green bars).  The top HCNN layer 7 best predicts IT responses. \textbf{(b.)} Similar to \textbf{a.}, but for macaque V4 neurons. Note that intermediate model layer 5 best predicts V4 responses.\textbf{(c.)} Representational similarity between visual representations in HCNN model layers and human V1-V3, based on fMRI data (adapted from~\cite{kriegeskorte:ploscb2014}).  Horizontal gray bar represents the inherent noise ceiling of the data.  Note that earlier HCNN model layers most resemble early visual areas.
\label{fig:results}}
\end{figure}

\begin{itemize}
	\item \emph{Inferior Temporal Cortex:}  Model responses from hidden layers near the top of HCNNs are highly predictive of neural responses in IT cortex, up to linear transform, both in electrophysiological~\cite{yamins:pnas2014, cadieu2014deep}, and fMRI data~\cite{kriegeskorte:ploscb2014, gucclu2015deep}. In published works such as Yamins et al 2014~\cite{yamins:pnas2014}, the predictivities are approximately 50\% explained variance (e.g. 70\% correlation similarity).  However, this number does not take into account the noise ceiling implied by inter-animal variability. Very recent work (personal communication) suggests that the inter-animal correlation similarity up to linear transform is about 0.8, and thus that the normalized correlation similarity is about 0.93, i.e. an explained variance of about 87\%. 
	
	Importantly for our purposes, the HCNN models are also substantially better at predicting neural response variance in IT than ideal-observer semantic models which have perfect access to object category or other attributes~\cite{yamins:pnas2014}. Though the ideal-observer models ``solve'' the posited behavioral task (e.g. categorization) perfectly, they are \emph{not} in fact runnable models at all, and so are not constrained to generate the answer from only the real inputs that the system has (e.g pixels). The fact that the HCNNs, which \emph{are} runnable models, end up being much better predictors of neural responses shows that runnability is an important constraint on the system. 

	\item \emph{Intermediate Cortical Areas:}  Intermediate layers of the same HCNNs whose later layers match IT neurons also yield state-of-the-art predictions of neural responses in V4 cortex~\cite{yamins:pnas2014,gucclu2015deep}, the dominant cortical input to IT.  Again, the mapping between models and brain data uses a linear transform to perform the match. Similarly, recent models with especially good task performance have distinct layers clearly segregating late-intermediate visual area PIT neurons from downstream central IT (CIT) and AIT neurons~\cite{nayebi2018task}.  These results are important because they show that high-level ecologically-relevant constraints on network function --- i.e. the categorization task imposed at the network's output layer --- are strong enough to shape upstream neural responses in a non-trivial way.  In other words, HCNN models suggest that the computations performed by the circuits V4 are structured \textit{in order that} that downstream computations in PIT and, subsequently, AIT, can support robust categorization in tasks that require the ability to deal with high-variation images.
	
   \item \emph{Early Visual Cortex:} Results in early visual cortex are equally striking. Extending the correspondence between HCNN layers and ventral stream layers down further, it has been shown that lower HCNN layers match neural responses in early visual cortex areas such as V1~\cite{kriegeskorte:ploscb2014, gucclu2015deep}. The filters that emerge from the learning process in early HCNN layers naturally resemble the Gabor wavelets seen qualitatively in V1~\cite{Carandini_2485}, without modelers having to build this structure in explicitly~\cite{Krizhevsky:2012wl}. In fact, recent high-resolution results show that early-intermediate layers of performance-optimized HCNNs are substantially better models of macaque V1 neural responses to natural images than previous state-of-the-art models that were hand-designed to replicate qualitative neuroscience observations such as those described in Fig 2b~\cite{cadena2019deep}. 
\end{itemize}

\noindent Taking all these results together, what has been shown is that a deep HCNN is a runnable model that can be mapped, at the spike-rate abstraction level, to both the structure and functional activity of the ventral visual pathway, using the same transform classes that are needed to map two animals' ventral pathways to each other.  In other words, the categorical aspects of the 3M++ criteria are satisfied. Moreover, the quantitative accuracy of the mapping can be assessed: as mentioned above, the match of HCNN activities to real neural activities in corresponding brain areas, using the linear transform similarity class, is reasonably quantitatively good (though not perfect).  

Before the deep HCNN results described above, there had not yet been direct experimental evidence that the neural responses throughout the visual system could be adequately characterized as a series of LN operations, organized into a feedforward hierarchical cascade, purely defined in terms of spike rates. Because these assumptions had not been experimentally confirmed, it has been suggested that to capture the (spike-rate) responses in downstream areas of the ventral pathway, it might be necessary for modelers to employ either primitives more complex than LN units~\cite{poirazi2003pyramidal}, or to introduce recurrent connections and long-range inter-area feedbacks~\cite{gilbert2013top}, or to model computations at a finer scale of abstraction than spike-rates, such as (e.g.) at the level of individual spikes~\cite{ghosh2009spiking}.
The existence of a \emph{runnable} feedforward LN-cascade model that reasonably predicts spike-rates throughout the ventral pathway gives an existence proof that these more complex hypotheses are largely not yet necessary, at least not for explaining the capacity to perform this particular sort of fast categorization task.

\textbf{The meaning of imperfection.} It is worth focusing for a moment on the imperfections of the HCNN models, and what implications those imperfections have for our advocacy of the 3M++ criteria.  It is clear that HCNNs do \emph{not} explain 100\% of the stimulus-driven variability of ventral pathway neurons, up to inter-animal noise ceiling. This failure is not just a mere trifle.  Finding a model that does meet the strict 100\%-explained-variance criterion would ultimately be needed to complete the mechanistic program required by 3M++.
Moreover, it is possible (and indeed likely) that some of the key structural issues described above (e.g. the lack of feedback and recurrent connections) will need to be resolved to develop models that map to the visual system with 100\% statistical accuracy, i.e. statistically indistinguishable from inter-animal maps.  

However -- and this is a crucial point for understanding our contribution in this paper -- such specific imperfections of the model class, and the changes needed to rectify them, will not substantively change our arguments that the 3M++ criteria provide a good framework for describing (and quantifying) what is required of a good mechanistic model in the first place.  Even as improved models (e.g. recurrent deep neural networks with long-range feedback connections) are developed, and the quantitative fits refined, the form of all the main ingredients -- the abstractions, the transform similarities, and how they fit together -- will remain the same.

\section{Using 3M++ to triangulate the level of abstraction }

To illustrate the significance of picking the right level of abstraction, consider a third, more loosely analogous case of digital computer chips. In such chips, there is a level of description at which it is best to think of the operation of the chip as digital. The circuit gates describing the system at this level are appropriately modelled with the abstract mathematics of Boolean logic, on which higher-level algorithmic abstractions can in turn be built. However, the underlying physical system implementing the chip is actually analog (since all real physics is analog).  The voltage equations governing the chip can in theory have much more complex response patterns than allowed by the digital logic gate that the circuit component is used to implement. If one these wanted to describe these further patterns with “digital logic” they’d be very much more complicated than the single simple gate that actually describes the component’s I/O.  Why is this?  The reason why digital programming works is that during normal usage the chip is clocked at a safe speed, i.e. driven by electrical switching events with a frequency below a pre-determined cutoff. In the safe regime, the complex voltage-driven response patterns of the analog guts of the circuit component can be guaranteed to produce a final I/O relationship that is consistent with the Boolean logic gate: ``digital discipline'' has been achieved.  It is possible to clock the system at a faster frequency than safe for maintaining digital discipline (this is called ``over-clocking'') but the correct operation of the chip can no longer be assumed.  Faster chips are produced by smaller and more precise components at which higher clock frequencies are possible without violating digital discipline. 

We are not suggesting that the situation with neurons is tightly connected to that of electronic circuits in any mechanistic way. However, at a high level the relationship between the analog chip implementation and its digital description is somewhat analogous to the relationship between the biophysics of the neuron and its LN-level spike-rate description. The lower-level descriptions are in both cases more complex, and when one tries to use the high-level description language (Boolean gates for CPUs, LN layers for neural networks) to describe the operations at a too-low level, the complexity of the description blows up as its fidelity to the physical mechanism falls apart. 
In the case of the 7-layer MLP for the L5PC cell discussed earlier, a deep neural network has been used merely as a nonlinear regression method --- definitely not useless as a data analysis tool, but not implementing what we would consider a predictively adequate runnable abstraction for the purposes of the 3M++ criteria. By contrast, in both the case of the CPU chip and the spike-rate-DNN, the coarser-grain descriptions do – unlike the seven-layer biophysical MLP – satisfy the 3M++ requirements when assessed at their proper level of abstraction.  

So it is important to emphasize that the results described in the previous sections indicated a good match between DNNs and neurons precisely because they did \textit{not} descend to the biophysical level. Instead of requiring a mapping to the ion channel and voltage level of description, the ``neurons'' of the DNNs in that work are mapped only to real neurons' spiking rates. While this means they make less detailed predictions, the spike-rate-DNNs are operating at a completely self-consistent description level (another way of saying that they have enough detail to actually be ``run''), and -- at their stipulated level of detail -- the spike-rate networks are both component-to-component mappable and predictively accurate.

Much of what we have said here may be reminiscent of Marr's influential distinction between the implementational level and the algorithmic. We think that Marr's terminology, while helpful in some cases, can lead us astray in others. This is because it conflates the value of providing abstract descriptions with the value of providing representational descriptions at an algorithmic level, descriptions all examples of which were \textit{also} both concise and intuitively interpretable.

As we have seen, in real biological systems, such concise, intermediate-level descriptions may not be available -- not just because we haven't found them yet, but because they may not \textit{exist}, given the nature of the system. Instead, we should understand that implementional facts (what we have been calling "mechanism", too, can come at different levels of abstraction. The way that these mechanistic facts produce the function or behavior (or "computation", as Marr called his top level) of interest, does not go via simple algorithms involving variables with clear psychological interpretations, or subfunctions whose composition intuitively produces the desired outcome, but, rather as we discuss in the companion paper,\footnote{See \url{https://arxiv.org/abs/2104.01489}.} are shaped by selection processes. We can probe those selection processes, but we cannot expect anything like Marr's algorithmic level to show up.

Respecting the 3M++ constraint forces us to build explanatory models whose parameters have clear interpretations as values of causal variables. Because the parameters have some significance not bound to the particular data-set being fit, they can generalize to new contexts. Models that satisfy 3M++ capture real causal patterns and dependencies in the world. 
  

\newpage

\pagenumbering{gobble}
\footnotesize{
\bibliographystyle{naturemag}
\linespread{0.9}
\bibliography{refs}

\begin{thebibliography}{10}
\expandafter\ifx\csname url\endcsname\relax
  \def\url#1{\texttt{#1}}\fi
\expandafter\ifx\csname urlprefix\endcsname\relax\def\urlprefix{URL }\fi
\providecommand{\bibinfo}[2]{#2}
\providecommand{\eprint}[2][]{\url{#2}}

\bibitem{machamer2000thinking}
\bibinfo{author}{Machamer, P.}, \bibinfo{author}{Darden, L.} \&
  \bibinfo{author}{Craver, C.~F.}
\newblock \bibinfo{title}{Thinking about mechanisms}.
\newblock \emph{\bibinfo{journal}{Philosophy of science}}
  \textbf{\bibinfo{volume}{67}}, \bibinfo{pages}{1--25} (\bibinfo{year}{2000}).

\bibitem{glennan1996mechanisms}
\bibinfo{author}{Glennan, S.~S.}
\newblock \bibinfo{title}{Mechanisms and the nature of causation}.
\newblock \emph{\bibinfo{journal}{Erkenntnis}} \textbf{\bibinfo{volume}{44}},
  \bibinfo{pages}{49--71} (\bibinfo{year}{1996}).

\bibitem{Craverbook2007}
\bibinfo{author}{Craver, C.~F.}
\newblock \emph{\bibinfo{title}{{Explaining the Brain: Mechanisms and the
  Mosaic Unity of Neuroscience}}} (\bibinfo{publisher}{Oxford University
  Press}, \bibinfo{year}{2007}).

\bibitem{levy2013abstraction}
\bibinfo{author}{Levy, A.} \& \bibinfo{author}{Bechtel, W.}
\newblock \bibinfo{title}{Abstraction and the organization of mechanisms}.
\newblock \emph{\bibinfo{journal}{Philosophy of science}}
  \textbf{\bibinfo{volume}{80}}, \bibinfo{pages}{241--261}
  (\bibinfo{year}{2013}).

\bibitem{10.5555/331969}
\bibinfo{author}{Pearl, J.}
\newblock \emph{\bibinfo{title}{Causality: Models, Reasoning, and Inference}}
  (\bibinfo{publisher}{Cambridge University Press}, \bibinfo{address}{USA},
  \bibinfo{year}{2000}).

\bibitem{pittphilsci10974}
\bibinfo{author}{Woodward, J.}
\newblock \bibinfo{title}{Explanation in neurobiology: An interventionist
  perspective}  (\bibinfo{year}{2014}).
\newblock \urlprefix\url{http://philsci-archive.pitt.edu/10974/}.

\bibitem{kaplan2011explanatory}
\bibinfo{author}{Kaplan, D.~M.} \& \bibinfo{author}{Craver, C.~F.}
\newblock \bibinfo{title}{The explanatory force of dynamical and mathematical
  models in neuroscience: A mechanistic perspective}.
\newblock \emph{\bibinfo{journal}{Philosophy of science}}
  \textbf{\bibinfo{volume}{78}}, \bibinfo{pages}{601--627}
  (\bibinfo{year}{2011}).

\bibitem{ross2015dynamical}
\bibinfo{author}{Ross, L.~N.}
\newblock \bibinfo{title}{Dynamical models and explanation in neuroscience}.
\newblock \emph{\bibinfo{journal}{Philosophy of Science}}
  \textbf{\bibinfo{volume}{82}}, \bibinfo{pages}{32--54}
  (\bibinfo{year}{2015}).

\bibitem{chirimuuta2017explanation}
\bibinfo{author}{Chirimuuta, M.}
\newblock \bibinfo{title}{Explanation in computational neuroscience: Causal and
  non-causal}.
\newblock \emph{\bibinfo{journal}{The British Journal for the Philosophy of
  Science}} \textbf{\bibinfo{volume}{69}}, \bibinfo{pages}{849--880}
  (\bibinfo{year}{2017}).

\bibitem{shepherd2015foundations}
\bibinfo{author}{Shepherd, G.~M.}
\newblock \emph{\bibinfo{title}{Foundations of the neuron doctrine}}
  (\bibinfo{publisher}{Oxford University Press}, \bibinfo{year}{2015}).

\bibitem{pillow2008spatio}
\bibinfo{author}{Pillow, J.~W.} \emph{et~al.}
\newblock \bibinfo{title}{Spatio-temporal correlations and visual signalling in
  a complete neuronal population}.
\newblock \emph{\bibinfo{journal}{Nature}} \textbf{\bibinfo{volume}{454}},
  \bibinfo{pages}{995--999} (\bibinfo{year}{2008}).

\bibitem{Carandini:2005kw}
\bibinfo{author}{Carandini, M.} \emph{et~al.}
\newblock \bibinfo{title}{{Do We Know What the Early Visual System Does?}}
\newblock \emph{\bibinfo{journal}{Journal of Neuroscience}}
  \textbf{\bibinfo{volume}{25}}, \bibinfo{pages}{10577--10597}
  (\bibinfo{year}{2005}).

\bibitem{poirazi2003pyramidal}
\bibinfo{author}{Poirazi, P.}, \bibinfo{author}{Brannon, T.} \&
  \bibinfo{author}{Mel, B.~W.}
\newblock \bibinfo{title}{Pyramidal neuron as two-layer neural network}.
\newblock \emph{\bibinfo{journal}{Neuron}} \textbf{\bibinfo{volume}{37}},
  \bibinfo{pages}{989--999} (\bibinfo{year}{2003}).

\bibitem{james1890principles}
\bibinfo{author}{James, W.}, \bibinfo{author}{Burkhardt, F.},
  \bibinfo{author}{Bowers, F.} \& \bibinfo{author}{Skrupskelis, I.~K.}
\newblock \emph{\bibinfo{title}{The principles of psychology}},
  vol.~\bibinfo{volume}{1} (\bibinfo{publisher}{Macmillan London},
  \bibinfo{year}{1890}).

\bibitem{Carandini_2485}
\bibinfo{author}{Carandini, M.} \emph{et~al.}
\newblock \bibinfo{title}{Do we know what the early visual system does?}
\newblock \emph{\bibinfo{journal}{J Neurosci}} \textbf{\bibinfo{volume}{25}},
  \bibinfo{pages}{10577--97} (\bibinfo{year}{2005}).

\bibitem{movshon1978spatial}
\bibinfo{author}{Movshon, J.~A.}, \bibinfo{author}{Thompson, I.~D.} \&
  \bibinfo{author}{Tolhurst, D.~J.}
\newblock \bibinfo{title}{Spatial summation in the receptive fields of simple
  cells in the cat's striate cortex.}
\newblock \emph{\bibinfo{journal}{The Journal of physiology}}
  \textbf{\bibinfo{volume}{283}}, \bibinfo{pages}{53--77}
  (\bibinfo{year}{1978}).

\bibitem{Freeman:2011gl}
\bibinfo{author}{Freeman, J.} \& \bibinfo{author}{Simoncelli, E.}
\newblock \bibinfo{title}{Metamers of the ventral stream}.
\newblock \emph{\bibinfo{journal}{Nature Neuroscience}}
  \textbf{\bibinfo{volume}{14}}, \bibinfo{pages}{1195--1201}
  (\bibinfo{year}{2011}).

\bibitem{DiCarlo_2007}
\bibinfo{author}{DiCarlo, J.~J.} \& \bibinfo{author}{Cox, D.~D.}
\newblock \bibinfo{title}{Untangling invariant object recognition}.
\newblock \emph{\bibinfo{journal}{Trends Cogn Sci}}
  \textbf{\bibinfo{volume}{11}}, \bibinfo{pages}{333--41}
  (\bibinfo{year}{2007}).

\bibitem{DiCarlo_2012}
\bibinfo{author}{DiCarlo, J.~J.}, \bibinfo{author}{Zoccolan, D.} \&
  \bibinfo{author}{Rust, N.~C.}
\newblock \bibinfo{title}{How does the brain solve visual object recognition?}
\newblock \emph{\bibinfo{journal}{Neuron}} \textbf{\bibinfo{volume}{73}},
  \bibinfo{pages}{415--34} (\bibinfo{year}{2012}).

\bibitem{schmolesky_478}
\bibinfo{author}{Schmolesky, M.~T.} \emph{et~al.}
\newblock \bibinfo{title}{Signal timing across the macaque visual system}.
\newblock \emph{\bibinfo{journal}{J Neurophysiol}}
  \textbf{\bibinfo{volume}{79}}, \bibinfo{pages}{3272--8}
  (\bibinfo{year}{1998}).

\bibitem{Lennie_2463}
\bibinfo{author}{Lennie, P.} \& \bibinfo{author}{Movshon, J.~A.}
\newblock \bibinfo{title}{Coding of color and form in the geniculostriate
  visual pathway (invited review)}.
\newblock \emph{\bibinfo{journal}{J Opt Soc Am A Opt Image Sci Vis}}
  \textbf{\bibinfo{volume}{22}}, \bibinfo{pages}{2013--33}
  (\bibinfo{year}{2005}).

\bibitem{Schiller_49}
\bibinfo{author}{Schiller, P.}
\newblock \bibinfo{title}{Effect of lesion in visual cortical area v4 on the
  recognition of transformed objects}.
\newblock \emph{\bibinfo{journal}{Nature}} \textbf{\bibinfo{volume}{376}},
  \bibinfo{pages}{342--344} (\bibinfo{year}{1995}).

\bibitem{gallant1996neural}
\bibinfo{author}{Gallant, J.}, \bibinfo{author}{Connor, C.},
  \bibinfo{author}{Rakshit, S.}, \bibinfo{author}{Lewis, J.} \&
  \bibinfo{author}{Van~Essen, D.}
\newblock \bibinfo{title}{Neural responses to polar, hyperbolic, and cartesian
  gratings in area v4 of the macaque monkey}.
\newblock \emph{\bibinfo{journal}{Journal of Neurophysiology}}
  \textbf{\bibinfo{volume}{76}}, \bibinfo{pages}{2718--2739}
  (\bibinfo{year}{1996}).

\bibitem{Brincat_2008}
\bibinfo{author}{Brincat, S.~L.} \& \bibinfo{author}{Connor, C.~E.}
\newblock \bibinfo{title}{Underlying principles of visual shape selectivity in
  posterior inferotemporal cortex}.
\newblock \emph{\bibinfo{journal}{Nat Neurosci}} \textbf{\bibinfo{volume}{7}},
  \bibinfo{pages}{880--6} (\bibinfo{year}{2004}).

\bibitem{yau2012curvature}
\bibinfo{author}{Yau, J.~M.}, \bibinfo{author}{Pasupathy, A.},
  \bibinfo{author}{Brincat, S.~L.} \& \bibinfo{author}{Connor, C.~E.}
\newblock \bibinfo{title}{Curvature processing dynamics in macaque area v4}.
\newblock \emph{\bibinfo{journal}{Cerebral Cortex}} \bibinfo{pages}{bhs004}
  (\bibinfo{year}{2012}).

\bibitem{Hung:2005jh}
\bibinfo{author}{Hung, C.~P.}, \bibinfo{author}{Kreiman, G.},
  \bibinfo{author}{Poggio, T.} \& \bibinfo{author}{Dicarlo, J.~J.}
\newblock \bibinfo{title}{{Fast readout of object identity from macaque
  inferior temporal cortex}}.
\newblock \emph{\bibinfo{journal}{Science}} \textbf{\bibinfo{volume}{310}},
  \bibinfo{pages}{863--866} (\bibinfo{year}{2005}).

\bibitem{majaj2015simple}
\bibinfo{author}{Majaj, N.~J.}, \bibinfo{author}{Hong, H.},
  \bibinfo{author}{Solomon, E.~A.} \& \bibinfo{author}{DiCarlo, J.~J.}
\newblock \bibinfo{title}{Simple learned weighted sums of inferior temporal
  neuronal firing rates accurately predict human core object recognition
  performance}.
\newblock \emph{\bibinfo{journal}{The Journal of Neuroscience}}
  \textbf{\bibinfo{volume}{35}}, \bibinfo{pages}{13402--13418}
  (\bibinfo{year}{2015}).

\bibitem{rahnev_denison_2018}
\bibinfo{author}{Rahnev, D.} \& \bibinfo{author}{Denison, R.~N.}
\newblock \bibinfo{title}{Suboptimality in perceptual decision making}.
\newblock \emph{\bibinfo{journal}{Behavioral and Brain Sciences}}
  \textbf{\bibinfo{volume}{41}}, \bibinfo{pages}{e223} (\bibinfo{year}{2018}).

\bibitem{v4_fitting_reynolds_2012}
\bibinfo{author}{Sharpee, T.~O.}, \bibinfo{author}{Kouh, M.} \&
  \bibinfo{author}{Reyholds, J.~H.}
\newblock \bibinfo{title}{Trade-off between curvature tuning and position
  invariance in visual area v4.}
\newblock \emph{\bibinfo{journal}{PNAS}} \textbf{\bibinfo{volume}{110}},
  \bibinfo{pages}{11618--11623} (\bibinfo{year}{2012}).

\bibitem{hubel1962receptive}
\bibinfo{author}{Hubel, D.~H.} \& \bibinfo{author}{Wiesel, T.~N.}
\newblock \bibinfo{title}{Receptive fields, binocular interaction and
  functional architecture in the cat's visual cortex}.
\newblock \emph{\bibinfo{journal}{The Journal of physiology}}
  \textbf{\bibinfo{volume}{160}}, \bibinfo{pages}{106--154}
  (\bibinfo{year}{1962}).

\bibitem{pasupathy2002population}
\bibinfo{author}{Pasupathy, A.} \& \bibinfo{author}{Connor, C.}
\newblock \bibinfo{title}{Population coding of shape in area v4}.
\newblock \emph{\bibinfo{journal}{Nature neuroscience}}
  \textbf{\bibinfo{volume}{5}}, \bibinfo{pages}{1332--1338}
  (\bibinfo{year}{2002}).

\bibitem{dicarlo:tics_2007}
\bibinfo{author}{DiCarlo, J.~J.} \& \bibinfo{author}{Cox, D.~D.}
\newblock \bibinfo{title}{{Untangling invariant object recognition}}.
\newblock \emph{\bibinfo{journal}{Trends in Cognitive Sciences}}
  \textbf{\bibinfo{volume}{11}}, \bibinfo{pages}{333--341}
  (\bibinfo{year}{2007}).

\bibitem{hubel1959receptive}
\bibinfo{author}{Hubel, D.~H.} \& \bibinfo{author}{Wiesel, T.~N.}
\newblock \bibinfo{title}{Receptive fields of single neurones in the cat's
  striate cortex}.
\newblock \emph{\bibinfo{journal}{The Journal of physiology}}
  \textbf{\bibinfo{volume}{148}}, \bibinfo{pages}{574--591}
  (\bibinfo{year}{1959}).

\bibitem{ringach2002orientation}
\bibinfo{author}{Ringach, D.~L.}, \bibinfo{author}{Shapley, R.~M.} \&
  \bibinfo{author}{Hawken, M.~J.}
\newblock \bibinfo{title}{Orientation selectivity in macaque v1: diversity and
  laminar dependence}.
\newblock \emph{\bibinfo{journal}{Journal of Neuroscience}}
  \textbf{\bibinfo{volume}{22}}, \bibinfo{pages}{5639--5651}
  (\bibinfo{year}{2002}).

\bibitem{willmore2008berkeley}
\bibinfo{author}{Willmore, B.}, \bibinfo{author}{Prenger, R.~J.},
  \bibinfo{author}{Wu, M. C.-K.} \& \bibinfo{author}{Gallant, J.~L.}
\newblock \bibinfo{title}{The berkeley wavelet transform: a biologically
  inspired orthogonal wavelet transform}.
\newblock \emph{\bibinfo{journal}{Neural computation}}
  \textbf{\bibinfo{volume}{20}}, \bibinfo{pages}{1537--1564}
  (\bibinfo{year}{2008}).

\bibitem{cadena2019deep}
\bibinfo{author}{Cadena, S.~A.} \emph{et~al.}
\newblock \bibinfo{title}{Deep convolutional models improve predictions of
  macaque v1 responses to natural images}.
\newblock \emph{\bibinfo{journal}{PLoS computational biology}}
  \textbf{\bibinfo{volume}{15}}, \bibinfo{pages}{e1006897}
  (\bibinfo{year}{2019}).

\bibitem{gilbert2013top}
\bibinfo{author}{Gilbert, C.~D.} \& \bibinfo{author}{Li, W.}
\newblock \bibinfo{title}{Top-down influences on visual processing}.
\newblock \emph{\bibinfo{journal}{Nature Reviews Neuroscience}}
  \textbf{\bibinfo{volume}{14}}, \bibinfo{pages}{350} (\bibinfo{year}{2013}).

\bibitem{rajalingham2018large}
\bibinfo{author}{Rajalingham, R.} \emph{et~al.}
\newblock \bibinfo{title}{Large-scale, high-resolution comparison of the core
  visual object recognition behavior of humans, monkeys, and state-of-the-art
  deep artificial neural networks}.
\newblock \emph{\bibinfo{journal}{Journal of Neuroscience}}
  \textbf{\bibinfo{volume}{38}}, \bibinfo{pages}{7255--7269}
  (\bibinfo{year}{2018}).

\bibitem{craver2018moredetails}
\bibinfo{author}{Craver, C.~F.} \& \bibinfo{author}{Kaplan, D.~M.}
\newblock \bibinfo{title}{{Are More Details Better? On the Norms of
  Completeness for Mechanistic Explanations}}.
\newblock \emph{\bibinfo{journal}{The British Journal for the Philosophy of
  Science}} \textbf{\bibinfo{volume}{71}}, \bibinfo{pages}{287--319}
  (\bibinfo{year}{2018}).
\newblock \urlprefix\url{https://doi.org/10.1093/bjps/axy015}.
\newblock
  \eprint{https://academic.oup.com/bjps/article-pdf/71/1/287/32567818/axy015.pdf}.

\bibitem{bullmore2009complex}
\bibinfo{author}{Bullmore, E.} \& \bibinfo{author}{Sporns, O.}
\newblock \bibinfo{title}{Complex brain networks: graph theoretical analysis of
  structural and functional systems}.
\newblock \emph{\bibinfo{journal}{Nature reviews neuroscience}}
  \textbf{\bibinfo{volume}{10}}, \bibinfo{pages}{186--198}
  (\bibinfo{year}{2009}).

\bibitem{Stinson}
\bibinfo{author}{Stinson, C.}
\newblock \bibinfo{title}{From implausible artificial neurons to idealized
  cognitive models: Rebooting philosophy of artificial intelligence}
  (\bibinfo{year}{2019}).
\newblock \urlprefix\url{http://philsci-archive.pitt.edu/16602/}.
\newblock \bibinfo{note}{Forthcoming in Philosophy of Science}.

\bibitem{Boone2016}
\bibinfo{author}{Boone, W.} \& \bibinfo{author}{Piccinini, G.}
\newblock \bibinfo{title}{The cognitive neuroscience revolution}.
\newblock \emph{\bibinfo{journal}{Synthese}} \textbf{\bibinfo{volume}{193}},
  \bibinfo{pages}{1509--1534} (\bibinfo{year}{2016}).
\newblock \urlprefix\url{https://doi.org/10.1007/s11229-015-0783-4}.

\bibitem{Piccinini2011}
\bibinfo{author}{Piccinini, G.} \& \bibinfo{author}{Craver, C.}
\newblock \bibinfo{title}{Integrating psychology and neuroscience: functional
  analyses as mechanism sketches}.
\newblock \emph{\bibinfo{journal}{Synthese}} \textbf{\bibinfo{volume}{183}},
  \bibinfo{pages}{283--311} (\bibinfo{year}{2011}).
\newblock \urlprefix\url{https://doi.org/10.1007/s11229-011-9898-4}.

\bibitem{weisberg2012simulation}
\bibinfo{author}{Weisberg, M.}
\newblock \emph{\bibinfo{title}{Simulation and similarity: Using models to
  understand the world}} (\bibinfo{publisher}{Oxford University Press},
  \bibinfo{year}{2012}).

\bibitem{ocko2018emergent}
\bibinfo{author}{Ocko, S.~A.}, \bibinfo{author}{Hardcastle, K.},
  \bibinfo{author}{Giocomo, L.~M.} \& \bibinfo{author}{Ganguli, S.}
\newblock \bibinfo{title}{Emergent elasticity in the neural code for space}.
\newblock \emph{\bibinfo{journal}{Proceedings of the National Academy of
  Sciences}} \textbf{\bibinfo{volume}{115}}, \bibinfo{pages}{E11798--E11806}
  (\bibinfo{year}{2018}).

\bibitem{Kriegeskorte_2526}
\bibinfo{author}{Kriegeskorte, N.} \emph{et~al.}
\newblock \bibinfo{title}{Matching categorical object representations in
  inferior temporal cortex of man and monkey}.
\newblock \emph{\bibinfo{journal}{Neuron}} \textbf{\bibinfo{volume}{60}},
  \bibinfo{pages}{1126--41} (\bibinfo{year}{2008}).

\bibitem{schwarzkopf2011surface}
\bibinfo{author}{Schwarzkopf, D.~S.}, \bibinfo{author}{Song, C.} \&
  \bibinfo{author}{Rees, G.}
\newblock \bibinfo{title}{The surface area of human v1 predicts the subjective
  experience of object size}.
\newblock \emph{\bibinfo{journal}{Nature neuroscience}}
  \textbf{\bibinfo{volume}{14}}, \bibinfo{pages}{28--30}
  (\bibinfo{year}{2011}).

\bibitem{Downing:2001dr}
\bibinfo{author}{Downing, P.}, \bibinfo{author}{Jiang, Y.},
  \bibinfo{author}{Shuman, M.} \& \bibinfo{author}{Kanwisher, N.}
\newblock \bibinfo{title}{A cortical area selective for visual processing of
  the human body}.
\newblock \emph{\bibinfo{journal}{Science}} \textbf{\bibinfo{volume}{293}},
  \bibinfo{pages}{2470--2473} (\bibinfo{year}{2001}).

\bibitem{Kanwisher_546}
\bibinfo{author}{Kanwisher, N.}, \bibinfo{author}{McDermott, J.} \&
  \bibinfo{author}{Chun, M.~M.}
\newblock \bibinfo{title}{The fusiform face area: a module in human
  extrastriate cortex specialized for face perception}.
\newblock \emph{\bibinfo{journal}{J Neurosci}} \textbf{\bibinfo{volume}{17}},
  \bibinfo{pages}{4302--11} (\bibinfo{year}{1997}).

\bibitem{Epstein:1998uj}
\bibinfo{author}{Epstein, R.} \& \bibinfo{author}{Kanwisher, N.}
\newblock \bibinfo{title}{A cortical representation of the local visual
  environment}.
\newblock \emph{\bibinfo{journal}{Nature}} \textbf{\bibinfo{volume}{392}},
  \bibinfo{pages}{592--601} (\bibinfo{year}{1998}).

\bibitem{cybenko1989approximation}
\bibinfo{author}{Cybenko, G.}
\newblock \bibinfo{title}{Approximation by superpositions of a sigmoidal
  function}.
\newblock \emph{\bibinfo{journal}{Mathematics of control, signals and systems}}
  \textbf{\bibinfo{volume}{2}}, \bibinfo{pages}{303--314}
  (\bibinfo{year}{1989}).

\bibitem{poggio2017and}
\bibinfo{author}{Poggio, T.}, \bibinfo{author}{Mhaskar, H.},
  \bibinfo{author}{Rosasco, L.}, \bibinfo{author}{Miranda, B.} \&
  \bibinfo{author}{Liao, Q.}
\newblock \bibinfo{title}{Why and when can deep-but not shallow-networks avoid
  the curse of dimensionality: a review}.
\newblock \emph{\bibinfo{journal}{International Journal of Automation and
  Computing}} \textbf{\bibinfo{volume}{14}}, \bibinfo{pages}{503--519}
  (\bibinfo{year}{2017}).

\bibitem{wei2019protein}
\bibinfo{author}{Wei, G.-W.}
\newblock \bibinfo{title}{Protein structure prediction beyond alphafold}.
\newblock \emph{\bibinfo{journal}{Nature Machine Intelligence}}
  \textbf{\bibinfo{volume}{1}}, \bibinfo{pages}{336--337}
  (\bibinfo{year}{2019}).

\bibitem{hay2011models}
\bibinfo{author}{Hay, E.}, \bibinfo{author}{Hill, S.},
  \bibinfo{author}{Sch{\"u}rmann, F.}, \bibinfo{author}{Markram, H.} \&
  \bibinfo{author}{Segev, I.}
\newblock \bibinfo{title}{Models of neocortical layer 5b pyramidal cells
  capturing a wide range of dendritic and perisomatic active properties}.
\newblock \emph{\bibinfo{journal}{PLoS computational biology}}
  \textbf{\bibinfo{volume}{7}}, \bibinfo{pages}{e1002107}
  (\bibinfo{year}{2011}).

\bibitem{david2019single}
\bibinfo{author}{David, B.}, \bibinfo{author}{Idan, S.} \&
  \bibinfo{author}{Michael, L.}
\newblock \bibinfo{title}{Single cortical neurons as deep artificial neural
  networks}.
\newblock \emph{\bibinfo{journal}{bioRxiv}} \bibinfo{pages}{613141}
  (\bibinfo{year}{2019}).

\bibitem{fukushima1980neocognitron}
\bibinfo{author}{Fukushima, K.}
\newblock \bibinfo{title}{{Neocognitron: A self-organizing neural network model
  for a mechanism of pattern recognition unaffected by shift in position}}.
\newblock \emph{\bibinfo{journal}{Biol Cybernetics}}  (\bibinfo{year}{1980}).

\bibitem{Deng_3067}
\bibinfo{author}{Deng, J.}, \bibinfo{author}{Li, K.}, \bibinfo{author}{Do, M.},
  \bibinfo{author}{Su, H.} \& \bibinfo{author}{Fei-Fei, L.}
\newblock \bibinfo{title}{Construction and analysis of a large scale image
  ontology}.
\newblock In \emph{\bibinfo{booktitle}{Vision Sciences Society}}
  (\bibinfo{year}{2009}).

\bibitem{yamins:pnas2014}
\bibinfo{author}{Yamins*, D.} \emph{et~al.}
\newblock \bibinfo{title}{Performance-optimized hierarchical models predict
  neural responses in higher visual cortex}.
\newblock \emph{\bibinfo{journal}{Proceedings of the National Academy of
  Sciences}}  (\bibinfo{year}{2014}).

\bibitem{kriegeskorte:ploscb2014}
\bibinfo{author}{Khaligh-Razavi, S.~M.} \& \bibinfo{author}{Kriegeskorte, N.}
\newblock \bibinfo{title}{Deep supervised, but not unsupervised, models may
  explain it cortical representation}.
\newblock \emph{\bibinfo{journal}{PLOS Comp. Bio.}}  (\bibinfo{year}{2014}).

\bibitem{cadieu2014deep}
\bibinfo{author}{Cadieu, C.~F.} \emph{et~al.}
\newblock \bibinfo{title}{Deep neural networks rival the representation of
  primate it cortex for core visual object recognition}.
\newblock \emph{\bibinfo{journal}{PLoS computational biology}}
  \textbf{\bibinfo{volume}{10}}, \bibinfo{pages}{e1003963}
  (\bibinfo{year}{2014}).

\bibitem{gucclu2015deep}
\bibinfo{author}{G{\"u}{\c{c}}l{\"u}, U.} \& \bibinfo{author}{van Gerven,
  M.~A.}
\newblock \bibinfo{title}{Deep neural networks reveal a gradient in the
  complexity of neural representations across the ventral stream}.
\newblock \emph{\bibinfo{journal}{The Journal of Neuroscience}}
  \textbf{\bibinfo{volume}{35}}, \bibinfo{pages}{10005--10014}
  (\bibinfo{year}{2015}).

\bibitem{nayebi2018task}
\bibinfo{author}{Nayebi, A.} \emph{et~al.}
\newblock \bibinfo{title}{Task-driven convolutional recurrent models of the
  visual system}.
\newblock In \emph{\bibinfo{booktitle}{Advances in Neural Information
  Processing Systems}}, \bibinfo{pages}{5290--5301} (\bibinfo{year}{2018}).

\bibitem{Krizhevsky:2012wl}
\bibinfo{author}{Krizhevsky, A.}, \bibinfo{author}{Sutskever, I.} \&
  \bibinfo{author}{Hinton, G.}
\newblock \bibinfo{title}{{ImageNet classification with deep convolutional
  neural networks}}.
\newblock \emph{\bibinfo{journal}{Advances in Neural Information Processing
  Systems}}  (\bibinfo{year}{2012}).

\bibitem{ghosh2009spiking}
\bibinfo{author}{Ghosh-Dastidar, S.} \& \bibinfo{author}{Adeli, H.}
\newblock \bibinfo{title}{Spiking neural networks}.
\newblock \emph{\bibinfo{journal}{International journal of neural systems}}
  \textbf{\bibinfo{volume}{19}}, \bibinfo{pages}{295--308}
  (\bibinfo{year}{2009}).

\end{thebibliography}
}

\end{document}